\documentclass[aps,prd,amsmath,twocolumn,showpacs,nofootinbib]{revtex4-1}

\usepackage{graphicx}
\usepackage{hyperref}
\usepackage{color}

\hypersetup{colorlinks=true}
\pdfoutput=1

\def\be{\begin{equation}}
\def\ee{\end{equation}}
\def\bea{\begin{eqnarray}}
\def\eea{\end{eqnarray}}

\begin{document}

\title{A characterization of 3+1 spacetimes via the Simon-Mars tensor}

\author{Claire Som\'e}
\email{claire.some@obspm.fr}
\affiliation{
Laboratoire Univers et Th\'eories, UMR 8102 du CNRS,
Observatoire de Paris, Universit\'e Paris Diderot, F-92190 Meudon, France}
\author{Philippe Grandcl\'ement}
\email{philippe.grandclement@obspm.fr}
\affiliation{
Laboratoire Univers et Th\'eories, UMR 8102 du CNRS,
Observatoire de Paris, Universit\'e Paris Diderot, F-92190 Meudon, France}
\author{Eric Gourgoulhon}
\email{eric.gourgoulhon@obspm.fr}
\affiliation{
Laboratoire Univers et Th\'eories, UMR 8102 du CNRS,
Observatoire de Paris, Universit\'e Paris Diderot, F-92190 Meudon, France}

\date{19 December 2014}  

\begin{abstract} 
We present the 3+1 decomposition of the Simon-Mars tensor, which has
the property of being identically zero for a vacuum and asymptotically
flat spacetime if and only if the latter is locally isometric to the
Kerr spacetime. Using this decomposition we form two dimensionless
scalar fields. Computing these scalars provides a simple way of comparing
locally a generic (even non vacuum and non analytic) stationary spacetime
to Kerr. As an illustration, we evaluate the Simon-Mars scalars for
numerical solutions of the Einstein equations generated by boson stars
and neutron stars, for analytic solutions of the Einstein equations
such as Curzon-Chazy spacetime and $\delta=2$ Tomimatsu-Sato spacetime,
and for an approximate solution of the Einstein equations : the modified
Kerr metric, which is an example of a parametric deviation from Kerr
spacetime.
\end{abstract} 

\pacs{04.20.Ex, 04.25.dg, 04.20.Jb, 95.30.Sf}

\maketitle

%%%%%%%%%%%%%%%%%%%%%%%%%%%%%%%%%%%%%%%%%%%%%%%%%%%%%%%%%%%%%%%%%%%%%%%%%%%%%%%

\section{Introduction}

The Kerr metric \cite{Kerr63} is an exact solution of the Einstein
equations describing rotating black holes. It is generally accepted
that the compact object at our Galactic Center SgrA{*} is described
by this geometry \cite{Gen}. But, alternative asymptotically flat
compact objects are also studied in the literature as possible models
for SgrA{*} (a recent example is rotating boson stars \cite{GSG14,Rev}).
It could be interesting to have a mathematical tool measuring the
deviation from these spacetimes with respect to the Kerr one. To achieve
this goal, the Simon-Mars tensor has been chosen.\\

The Simon-Mars tensor has been introduced by Mars in \cite{Mars99},
under the name of Spacetime Simon tensor. This tensor and its ancestors,
the Cotton tensor \cite{Cotton} and the Simon tensor \cite{Si},
were defined to understand what singles out Kerr among the family
of stationary and axisymmetric metrics. In \cite{Mars99}, Mars proved
a theorem stating that if a spacetime satisfies the Einstein vacuum
field equations, is asymptotically flat and admit a smooth Killing
vector field $\vec{\xi}$ such that the Simon-Mars tensor associated
to $\vec{\xi}$ vanishes everywhere, then this spacetime is locally
isometric to a Kerr spacetime. In this work, we use this property
to quantify the ``non-Kerness'' of a given spacetime.\\

The name Simon-Mars tensor has already been used in \cite{Bini01}
for a 2-tensor linked by a duality relation (see Section \ref{sub:Definition})
to the 3-tensor used in this paper, so we keep the name of Simon-Mars
tensor. One has to be careful not to confuse this tensor with the
Mars-Simon tensor, defined in \cite{IK}, which is a 4-tensor, and
which can exist only in vacuum spacetimes, where an Ernst potential
can be defined (see Section \ref{sub:Definition}). The use of the
Simon-Mars tensor allows us to consider generic stationary spacetimes
which have a matter content, such as boson stars or neutron stars,
while it would not have any meaning for the Mars-Simon tensor. Nevertheless,
the 3+1 decomposition of the Mars-Simon tensor done in \cite{Kroon08}
presents some similarities with this work, so they are exploited in
this paper.\\

The object of this paper is to present the 3+1 decomposition of the
Simon-Mars tensor. The main motivation is to characterize numerical
spacetimes, which are usually given in 3+1 form. We use the theorem
cited above to quantify the ``non-Kerness'' of a given stationary
spacetime by seeing how much the Simon-Mars components of this spacetime
differ from zero. To work with coordinate independent quantities,
we form 2 scalar fields from the Simon-Mars tensor and we also give
their 3+1 decomposition.\\

The search for invariant quantities quantifying the ``non-Kerness''
of a metric can be used also for studying the stability of Kerr spacetime
under non-linear perturbations, which remains an open problem. Such
invariant have been built with tensors but only in vacuum spacetimes,
as in \cite{GLS13}, and also with spinors, as in \cite{Kroon05}.
The calculations with spinors are useful from a theoretical point
of view but very difficult to treat numerically.\\

In the following section, the definition and properties of the Simon-Mars
tensor are briefly reviewed. In Section \ref{sec:Orthogonal-splitting-of},
the 3+1 decomposition of this tensor in 8 components is performed
step by step. Then the definition of the Simon-Mars scalar fields
is given. In the next section (Section \ref{sec:Axisymmetric-spacetimes}),
the specific case of axisymmetric spacetimes is considered. The last
part, Sections \ref{sec:Numerical-applications} is devoted to the
applications : first we present numerical computation of the 3+1 components
of the Simon-Mars for a Kerr spacetime to verify that all of them
are identically zero. Then we compute those 8 components in adapted
coordinates for two numerical solutions of Einstein equations : rotating
boson stars and neutron stars, and we do the same for the 2 scalars.
To do the link between our characterization and the one of \cite{GLS13},
we tackle two examples given in \cite{GLS13} : Curzon-Chazy and $\delta=2$
Tomimatsu-Sato spacetimes, which are exact analytic solutions of the
Einstein equations. Finally, we compute the Simon-Mars scalars for
a family of spacetimes deviating from Kerr by a continuous parameter
: the modified Kerr metric \cite{JoPs}.

\section{Simon-Mars tensor}

\subsection{\label{sub:Definition}Definition}

Let $\mathcal{M}$ be a 4-dimensional manifold endowed with a smooth
Lorentzian metric $g_{\alpha\beta}$ of signature $\left(-,+,+,+\right)$.
Greek indices vary form 0 to 4 while Latin indices are only 1 to 3.
The Einstein summation convention is used unless specified otherwise.
We work in geometric units for which $G=c=1$. The Levi-Civita covariant
derivative associated to $g_{\alpha\beta}$ is the operator $\nabla_{\alpha}$
while $R_{\:\nu\rho\sigma}^{\mu}$ and $R_{\nu\sigma}$ denote respectively
the Riemann and Ricci tensors. From these two tensors one builds the
Weyl tensor $C_{\alpha\beta\gamma\delta}$ which corresponds to the
traceless part of the Riemann tensor%
\footnote{The Weyl tensor coincides with the Riemann tensor for vacuum spacetimes
such as Kerr spacetime.%
} : 
\begin{eqnarray}
C_{\alpha\beta\gamma\delta} & = & R_{\alpha\beta\gamma\delta}+\frac{R}{6}\left(g_{\alpha\gamma}g_{\beta\delta}-g_{\alpha\delta}g_{\beta\gamma}\right)\nonumber \\
 &  & -\frac{1}{2}\left(g_{\alpha\gamma}R_{\beta\delta}-g_{\alpha\delta}R_{\beta\gamma}-g_{\beta\gamma}R_{\alpha\delta}+g_{\beta\delta}R_{\alpha\gamma}\right).\label{eq:Wt-1}
\end{eqnarray}
Furthermore we use the right self-dual Weyl tensor, which is defined
as 
\begin{eqnarray}
\mathcal{C}_{\alpha\beta\gamma\delta} & = & C_{\alpha\beta\gamma\delta}+iC_{\alpha\beta\gamma\delta}^{\star}\nonumber \\
 & = & C_{\alpha\beta\gamma\delta}+\frac{1}{2}i\eta_{\gamma\delta\rho\sigma}C_{\alpha\beta}^{\:\:\:\rho\sigma},\label{eq:Csd-1}
\end{eqnarray}
where $\eta_{\gamma\delta\rho\sigma}$ is the volume 4-form associated
with $g$ and the star denotes Hodge duality : the Hodge dual of a
2-form is given by 
\begin{equation}
F_{\mu\nu}^{\star}=\frac{1}{2}\eta_{\mu\nu\lambda\rho}F^{\lambda\rho}.\label{eq:sd}
\end{equation}
To construct the Simon-Mars tensor one assumes the existence of a
Killing vector field in spacetime. Let us recall that a vector field
$\xi^{\mu}$ on $\mathcal{M}$ is called a Killing vector field if
it verifies the following condition : 
\begin{equation}
\mathcal{L}_{\xi}g_{\alpha\beta}=0,\label{eq:Lg}
\end{equation}
where $\mathcal{L}_{\xi}$ denotes the Lie derivative with respect
to $\xi$. As we are interested in this work by stationary spacetimes,
they all possess the Killing vector field $\xi=\partial_{t}$. Besides,
a Killing vector field satisfies the identity 
\begin{equation}
\nabla_{\left(\alpha\right.}\xi_{\left.\beta\right)}=0.\label{eq:Keq-1}
\end{equation}
To this property follow the definition of the Papapetrou field given
by 
\begin{equation}
F_{\alpha\beta}=\nabla_{\alpha}\xi_{\beta}.\label{eq:Fab}
\end{equation}
$F_{\alpha\beta}$ is antisymmetric (i.e. is a 2-form) because of
(\ref{eq:Keq-1}). We will also use the self-dual form of the Papapetrou
field defined as we have seen in (\ref{eq:Csd-1}) and (\ref{eq:Keq-1})
by 
\begin{equation}
\mathcal{F}_{\alpha\beta}=F_{\alpha\beta}+iF_{\alpha\beta}^{\star}=\nabla_{\alpha}\xi_{\beta}+\frac{i}{2}\eta_{\alpha\beta\lambda\mu}\nabla^{\lambda}\xi^{\mu}.\label{eq:sdFab}
\end{equation}
From $F_{\alpha\beta}^{\star}$ we define the twist 1-form 
\begin{equation}
\omega_{\alpha}=F_{\alpha\beta}^{\star}\xi^{\beta},\label{eq:t1}
\end{equation}
which is closed for a vacuum spacetime. In such a case we can define
a local potential which is called the twist potential $\omega$ :
$\omega_{\alpha}=\nabla_{\alpha}\omega$. The norm of the Killing
vector field is $\lambda=\xi_{\rho}\xi^{\rho}$ and no restriction
is imposed on its sign. The last ingredient needed to construct the
Simon-Mars tensor is the Ernst 1-form : 
\begin{equation}
\sigma_{\mu}=2\xi^{\alpha}\mathcal{F}_{\alpha\mu}.\label{eq:E1}
\end{equation}
In vacuum spacetimes, the Ernst 1-form is closed so one can define
a local potential $\sigma$ called the Ernst potential which can be
written in terms of the norm and twist of the Killing vector field
\begin{equation}
\sigma=\lambda+2i\omega,\label{eq:Ernst}
\end{equation}
but this is not the case for non-vacuum spacetimes such as boson star
and neutron stars spacetimes. At last we can give the definition of
the Simon-Mars tensor, given by Mars \cite{Mars99}, and based on the work of Simon \cite{Si} :
\begin{equation}
S_{\alpha\beta\nu}=4\xi^{\mu}\xi^{\rho}\mathcal{C}_{\mu\alpha\rho\left[\beta\right.}\sigma_{\left.\nu\right]}+\gamma_{\alpha\left[\beta\right.}\mathcal{C}_{\left.\nu\right]\mu\rho\delta}\mathcal{F}^{\rho\delta}\xi^{\mu},\label{eq:S}
\end{equation}
where we used the following abbreviation
\begin{equation}
\gamma_{\alpha\beta}=-\lambda g_{\alpha\beta}+\xi_{\alpha}\xi_{\beta}.\label{eq:gamma}
\end{equation}
As stated in the introduction, Bini et al. \cite{Bini01} are calling
Simon-Mars tensor the 2-tensor $\tilde{S}$ linked to $S$ by :
\begin{equation}
\tilde{S}_{\alpha\beta}=n^{\sigma}\eta_{\alpha\sigma\lambda\mu}S_{\beta}^{\;\lambda\mu}.\label{eq:SM2}
\end{equation}

\subsection{Properties}

The Simon-Mars tensor (\ref{eq:S}) has the algebraic properties of
a Lanczos potential, namely :
\begin{eqnarray}
S_{\;\beta\nu}^{\beta} & = & 0\nonumber \\
S_{\alpha\left[\beta\nu\right]} & = & S_{\alpha\beta\nu}\nonumber \\
S_{\alpha\beta\nu}+S_{\beta\nu\alpha}+S_{\nu\alpha\beta} & = & 0.\label{eq:Sprop}
\end{eqnarray}
But the fundamental property of this tensor used in this work and
derived in \cite{Mars99} is the following : the Simon-Mars tensor
$S_{\alpha\beta\nu}$ vanishes identically for an asymptotically flat
spacetime which verifies the Einstein vacuum field equations if and
only if this spacetime is locally isometric to the Kerr spacetime.
A geometric interpretation of this statement is discussed in \cite{Mars00}.\\

In the sense of the preceding quoted theorem, the Simon-Mars tensor
characterizes the Kerr spacetime. It is then interesting to calculate
its value for other asymptotically flat and stationary spacetimes.

\section{\label{sec:Orthogonal-splitting-of}Orthogonal splitting of the Simon-Mars
tensor}

\subsection{Basis of 3+1 formalism and useful formulas}

In this article, we only consider globally hyperbolic spacetimes.
Such spacetimes admit a foliation by a one-parameter family of spacelike
hypersurfaces denoted by $\Sigma$. This is the 3+1 decomposition,
references on this formalism can be found in the literature, see for
instance \cite{Gourg12,Alcub08}. Here we review only the formulas which are
useful for this work. \\

The unit vector which determines the unique direction normal to $\Sigma$,
denoted by $n^{\alpha}$, is also the 4-velocity of an observer called
the Eulerian observer. Because $\Sigma$ is spacelike, the following
property is verified by the normal vector
\begin{equation}
n^{\alpha}n_{\alpha}=-1.\label{eq:nn}
\end{equation}
On each hypersurface the metric induced by $g$ is 
\begin{equation}
h_{\alpha\beta}=g_{\alpha\beta}+n_{\alpha}n_{\beta},\label{eq:met}
\end{equation}
and the extrinsic curvature is given by 
\begin{equation}
K_{\alpha\beta}=-\frac{1}{2}\mathcal{L}_{n}h_{\alpha\beta}.\label{eq:Kab}
\end{equation}
We will also use the following abbreviation
\begin{equation}
l_{\alpha\beta}=h_{\alpha\beta}+n_{\alpha}n_{\beta}=g_{\alpha\beta}+2n_{\alpha}n_{\beta}.\label{eq:l}
\end{equation}
The orthogonal splitting of the volume element is given by
\begin{eqnarray}
\eta_{\alpha\beta\gamma\delta} & = & -n_{\alpha}\epsilon_{\beta\gamma\delta}+n_{\beta}\epsilon_{\alpha\gamma\delta}\nonumber \\
 &  & -n_{\gamma}\epsilon_{\alpha\beta\delta}+n_{\delta}\epsilon_{\alpha\beta\gamma},\label{eq:As}
\end{eqnarray}
where $\epsilon_{\alpha\beta\gamma}=n^{\lambda}\eta_{\lambda\alpha\beta\gamma}$
is the spatial volume element which is a fully antisymmetric spatial
tensor and which verifies 
\begin{eqnarray}
\epsilon_{ijk}\epsilon^{ijl} & = & 2h_{k}^{\enskip l},\label{eq:EpsEps1}\\
\epsilon_{ijk}\epsilon^{lmn} & = & h_{i}^{\enskip l}h_{j}^{\enskip m}h_{k}^{\enskip n}+h_{i}^{\enskip m}h_{j}^{\enskip n}h_{k}^{\enskip l}\nonumber \\
 &  & +h_{i}^{\enskip n}h_{j}^{\enskip l}h_{k}^{\enskip m}-h_{i}^{\enskip l}h_{j}^{\enskip n}h_{k}^{\enskip m}\nonumber \\
 &  & -h_{i}^{\enskip m}h_{j}^{\enskip l}h_{k}^{\enskip n}-h_{i}^{\enskip n}h_{j}^{\enskip m}h_{k}^{\enskip l}.\label{eq:EpsEps2}
\end{eqnarray}

In all the calculation, we suppose the existence of the Killing vector
field $\xi^{\mu}=\partial_{t}^{\mu}$ (because all the spacetimes
considered are stationary). Its orthogonal splitting is given by :
\begin{equation}
\xi^{\mu}=\partial_{t}^{\mu}=Nn^{\mu}+\beta^{\mu},\label{eq:KV}
\end{equation}
where $N$ is called the lapse because it is related to the time lapse
between two slices of the foliation and $\beta^{\mu}$ the shift because
it tells how the coordinates are shifted from one slice to another,
cf \cite{Gourg12} for details. The line element of a spacetime expressed
with the 3+1 formalism is the following : 
\begin{eqnarray}
g_{\alpha\beta}\textrm{d}x^{\alpha}\textrm{d}x^{\beta} & = & -(N^{2}-h_{ij}\beta^{i}\beta^{j})\textrm{d}t^{2}+h_{ij}\textrm{d}x^{i}\textrm{d}x^{j}\nonumber \\
 &  & +2h_{ij}\beta^{i}\textrm{d}x^{j}\textrm{d}t.\label{eq:31met}
\end{eqnarray}
We end this section by writing, for the specific case of stationarity
(all the partial derivatives with respect to the time are zero), every
useful formula for the next section (each of them can be easily derived)
:
\begin{itemize}
\item The 3+1 decomposition of the derivative of the unit normal vector
: 
\begin{equation}
\nabla_{\alpha}n_{\beta}=-K_{\alpha\beta}-n_{\alpha}D_{\beta}\ln N,\label{eq:deln}
\end{equation}
where $D_{\alpha}$ is the covariant derivative associated with the
3-dimensional metric $h$.
\item The derivative of the lapse :
\begin{equation}
\nabla_{\alpha}N=D_{\alpha}N+n_{\alpha}\beta^{\gamma}D_{\gamma}\ln N.\label{eq:delN}
\end{equation}

\item The Lie derivative of the shift :
\end{itemize}
\begin{equation}
\mathcal{L}_{n}\beta_{\alpha}=-2\beta^{\gamma}K_{\alpha\gamma}.\label{eq:LB}
\end{equation}

\begin{itemize}
\item The derivative of the shift :
\begin{eqnarray}
\nabla_{\alpha}\beta_{\gamma} & = & D_{\alpha}\beta_{\gamma}-n_{\gamma}\beta_{\delta}K_{\alpha}^{\;\delta}\nonumber \\
 &  & -n_{\alpha}\left(n_{\gamma}\beta^{\delta}D_{\delta}\ln N-\beta^{\delta}K_{\delta\gamma}\right).\label{eq:delb}
\end{eqnarray}

\end{itemize}

\subsection{Orthogonal splitting of the self-dual Weyl tensor}

First we introduce the electric and magnetic parts of the Weyl tensor
given in \cite{Alcub08} (and first defined in \cite{Matte53}) 
\begin{eqnarray}
E_{\alpha\beta} & = & C_{\alpha\gamma\beta\delta}n^{\gamma}n^{\delta},\label{eq:E}\\
B_{\alpha\beta} & = & \frac{1}{2}\eta_{\beta\delta\rho\sigma}C_{\alpha\gamma}^{\quad\rho\sigma}n^{\gamma}n^{\delta}.\label{eq:B}
\end{eqnarray}
The decomposition of the Weyl tensor (\ref{eq:Wt-1}) using (\ref{eq:E})
and (\ref{eq:B}) is also given in \cite{Alcub08} using (\ref{eq:l})
(for a demonstration see \cite{KroonBook}) 
\begin{eqnarray}
C_{\alpha\beta\gamma\delta} & = & 2\left(l_{\alpha\left[\gamma\right.}E_{\left.\delta\right]\beta}+l_{\beta\left[\delta\right.}E_{\left.\gamma\right]\alpha}\right)\nonumber \\
 &  & -2\left(\epsilon_{\gamma\delta\rho}n_{\left[\alpha\right.}B_{\left.\beta\right]}^{\;\;\rho}+\epsilon_{\alpha\beta\rho}n_{\left[\gamma\right.}B_{\left.\delta\right]}^{\;\;\rho}\right),\label{eq:Wt2}
\end{eqnarray}
so (\ref{eq:E}) and (\ref{eq:B}) are given by (again see \cite{Alcub08})
\begin{eqnarray}
E_{ij} & = & R_{ij}+KK_{ij}-K_{il}K_{\enskip j}^{l}\nonumber \\
 &  & -4\pi\left[S_{ij}+\frac{h_{ij}}{3}\left(4\rho-S\right)\right],\label{eq:E31}\\
B_{ij} & = & \epsilon_{i}^{\enskip mn}\left(D_{m}K_{nj}-4\pi h_{jm}p_{n}\right),\label{eq:B31}
\end{eqnarray}
with $S_{\alpha\beta}$, $p_{\alpha}$ and $\rho$ the components
of the orthogonal splitting of the stress energy tensor $T_{\mu\nu}$
: 
\begin{eqnarray}
\rho & = & T_{\mu\nu}n^{\mu}n^{\nu},\label{eq:rho}\\
p_{\alpha} & = & -T_{\mu\nu}h_{\:\alpha}^{\mu}n^{\nu},\label{eq:pa}\\
S_{\alpha\beta} & = & T_{\mu\nu}h_{\:\alpha}^{\mu}h_{\:\beta}^{\nu}.\label{eq:Sab}
\end{eqnarray}
Thanks to (\ref{eq:Csd-1}), (\ref{eq:Wt2}) is also the 3+1 decomposition
of the real part of the self-dual Weyl tensor. As the magnetic part
of the Weyl tensor is the Hodge dual of the electric part, the imaginary
part of the self-dual Weyl tensor reads 
\begin{eqnarray}
\textrm{Im}\left(C_{\alpha\beta\gamma\delta}\right) & = & 2\left(l_{\beta\left[\gamma\right.}B_{\left.\delta\right]\alpha}+l_{\alpha\left[\delta\right.}B_{\left.\gamma\right]\beta}\right)\nonumber \\
 &  & +2i\left(\epsilon_{\delta\gamma\rho}n_{\left[\alpha\right.}E_{\left.\beta\right]}^{\;\;\rho}+\epsilon_{\alpha\beta\rho}n_{\left[\delta\right.}E_{\left.\gamma\right]}^{\;\;\rho}\right).\label{eq:ImC}
\end{eqnarray}

\subsection{Orthogonal splitting of $\mathcal{F}_{\alpha\beta}$, $\sigma_{\mu}$
and $\gamma_{\alpha\beta}$ }

The self-dual Papapetrou field is defined by (\ref{eq:sdFab}). As
$\mathcal{F}_{\alpha\beta}$ is complex, we perform the 3+1 decomposition
first for the real part and then for the imaginary one. The real part
is (\ref{eq:Fab}), so we take the 3+1 decomposition of the Killing
vector field $\xi_{\beta}$ given by (\ref{eq:KV}), we develop, then
we use (\ref{eq:deln}), (\ref{eq:delN}) and (\ref{eq:delb}). The
antisymmetric part reads%
\footnote{same as equation (4.35) in \cite{Kroon08}%
} 
\begin{equation}
\textrm{Re}\left(\mathcal{F}_{\alpha\beta}\right)=D_{\left[\alpha\right.}\beta_{\left.\beta\right]}-2n_{\left[\alpha\right.}D_{\left.\beta\right]}N+2n_{\left[\alpha\right.}\beta^{\delta}K_{\left.\beta\right]\delta}.\label{eq:ReF}
\end{equation}
We checked also that we recover the Killing equation (\ref{eq:Keq-1})
with the symmetric part. Let us do the same for the imaginary part,
after development and using the expression of the volume form (\ref{eq:As})
and its antisymmetry we obtain%
\footnote{except for the sign misprint in the second term, same as equation
(4.36) in \cite{Kroon08}%
} 
\begin{eqnarray}
\textrm{Im}\left(\mathcal{F}_{\alpha\beta}\right) & = & -n_{\left[\alpha\right.}\epsilon_{\left.\beta\right]\lambda\mu}D^{\lambda}\beta^{\mu}\nonumber \\
 &  & -\epsilon_{\alpha\beta\mu}\left(D^{\mu}N-\beta^{\delta}K_{\enskip\delta}^{\mu}\right).\label{eq:ImF}
\end{eqnarray}
Let us tackle the decomposition of the Ernst potential given by (\ref{eq:E1})
\begin{equation}
\sigma_{\mu}=2\xi^{\alpha}\nabla_{\alpha}\xi_{\mu}+i\xi^{\alpha}\eta_{\alpha\mu\lambda\nu}\nabla^{\lambda}\xi^{\nu}.\label{eq:E12}
\end{equation}
Using (\ref{eq:KV}) and (\ref{eq:ReF}) (resp. (\ref{eq:ImF})) the
3+1 decomposition of the real part (resp. imaginary part) is given
by%
\footnote{same as real and imaginary part of equation (6.5) in \cite{Kroon08},
except for one term forgotten in the imaginary part%
} 
\begin{eqnarray}
\textrm{Re}\left(\sigma_{\mu}\right) & = & 2ND_{\mu}N-2NK_{\mu i}\beta^{i}+2\beta^{i}D_{\left[i\right.}\beta_{\left.\mu\right]}\nonumber \\
 &  & +2n_{\mu}\left(\beta^{i}D_{i}N-\beta_{j}\beta^{i}K_{i}^{\; j}\right),\label{eq:Resig}\\
\textrm{Im}\left(\sigma_{\mu}\right) & = & \epsilon_{\mu ij}\left(2\beta^{i}D^{j}N-2K^{jk}\beta_{k}\beta^{i}+ND^{i}\beta^{j}\right)\nonumber \\
 &  & +n_{\mu}\epsilon_{ijk}\beta^{i}D^{j}\beta^{k}.\label{eq:Imsig}
\end{eqnarray}
Finally we can split $\gamma_{\alpha\beta}$ given by (\ref{eq:gamma})
simply using (\ref{eq:met}) and (\ref{eq:KV})
\begin{eqnarray}
\gamma_{\alpha\beta} & = & \left(N^{2}-\beta_{i}\beta^{i}\right)h_{\alpha\beta}+\beta_{\alpha}\beta_{\beta}\nonumber \\
 &  & +N\left(n_{\alpha}\beta_{\beta}+\beta_{\alpha}n_{\beta}\right)+\beta_{i}\beta^{i}n_{\alpha}n_{\beta}.\label{eq:gamma31}
\end{eqnarray}

\subsection{Orthogonal splitting of the Simon-Mars tensor}

Let us recall the definition of the Simon-Mars tensor (\ref{eq:S})
\[
S_{\alpha\beta\nu}=4\underset{FT}{\underbrace{\xi^{\mu}\xi^{\rho}\mathcal{C}_{\mu\alpha\rho\left[\beta\right.}\sigma_{\left.\nu\right]}}}+\underset{ST}{\underbrace{\gamma_{\alpha\left[\beta\right.}\mathcal{C}_{\left.\nu\right]\mu\rho\delta}\mathcal{F}^{\rho\delta}\xi^{\mu}}}.
\]
As this is a complex tensor, we decompose independently the real and
the imaginary parts, and we also consider one of the two terms of
(\ref{eq:S}) at a time. 

\subsubsection{First term }

Let us then first consider the fourth of the real part of the first
term and develop it 
\begin{eqnarray}
\textrm{Re}\left(FT\right) & = & \underset{I_{\alpha\beta\nu}}{\underbrace{\xi^{\mu}\xi^{\rho}\textrm{Re}\left(\mathcal{C}_{\mu\alpha\rho\left[\beta\right.}\right)\textrm{Re}\left(\sigma_{\left.\nu\right]}\right)}}\nonumber \\
 &  & -\underset{II_{\alpha\beta\nu}}{\underbrace{\xi^{\mu}\xi^{\rho}\textrm{Im}\left(\mathcal{C}_{\mu\alpha\rho\left[\beta\right.}\right)\textrm{Im}\left(\sigma_{\left.\nu\right]}\right)}}.\label{eq:FTRe}
\end{eqnarray}
Using (\ref{eq:KV}), (\ref{eq:Wt2}), we obtain the 3+1 decomposition
of the first part of $I_{\alpha\beta\nu}$, 
\begin{equation}
\xi^{\mu}\xi^{\rho}\textrm{Re}\left(\mathcal{C}_{\mu\alpha\rho\beta}\right)=V_{\alpha\beta}+n_{\alpha}W_{\beta}+n_{\beta}W_{\alpha}+n_{\alpha}n_{\beta}Y,\label{eq:KKReC}
\end{equation}
with
\begin{eqnarray}
V_{ij} & = & \left(N^{2}+\beta_{k}\beta^{k}\right)E_{ij}-\beta_{j}\beta^{k}E_{ik}-\beta^{k}\beta_{i}E_{kj}\nonumber \\
 &  & +h_{ij}\beta^{k}\beta^{\rho}E_{k\rho}+N\left(\epsilon_{kjs}\beta^{k}B_{i}^{\enskip s}+\epsilon_{kis}\beta^{k}B_{j}^{\enskip s}\right),\label{eq:Vab}\\
W_{i} & = & N\beta^{k}E_{ik}+\epsilon_{kis}\beta^{l}\beta^{k}B_{l}^{\enskip s},\label{eq:Wb}\\
Y & = & \beta^{i}\beta^{j}E_{ij}.\label{eq:Y}
\end{eqnarray}
Rewriting (\ref{eq:Resig}) we have directly the decomposition of
the second part 
\begin{equation}
\textrm{Re}\left(\sigma_{\nu}\right)=Z_{\nu}+n_{\nu}T,\label{eq:ReS}
\end{equation}
with
\begin{eqnarray}
Z_{i} & = & 2ND_{i}N-2NK_{ik}\beta^{k}+2\beta^{k}D_{\left[k\right.}\beta_{\left.i\right]},\label{eq:Zv}\\
T & = & 2\beta^{i}D_{i}N-2K_{ij}\beta^{i}\beta^{j},\label{eq:T}
\end{eqnarray}
so
\begin{eqnarray}
I_{\alpha\beta\nu} & = & 2V_{\alpha\left[\beta\right.}Z_{\left.\nu\right]}+2TV_{\alpha\left[\beta\right.}n_{\left.\nu\right]}\nonumber \\
 &  & +2n_{\alpha}W_{\left[\beta\right.}Z_{\left.\nu\right]}+2W_{\alpha}n_{\left[\beta\right.}Z_{\left.\nu\right]}\nonumber \\
 &  & +2TW_{\left[\beta\right.}n_{\left.\nu\right]}n_{\alpha}+2Yn_{\alpha}n_{\left[\beta\right.}Z_{\left.\nu\right]}.\label{eq:ReFTA}
\end{eqnarray}
Using (\ref{eq:ImC}), we do the same for the first part of $II_{\alpha\beta\nu}$
of (\ref{eq:FTRe}), we obtain
\begin{equation}
\xi^{\mu}\xi^{\rho}\textrm{Im}\left(\mathcal{C}_{\mu\alpha\rho\beta}\right)=\bar{V}_{\alpha\beta}+n_{\alpha}\bar{W}_{\beta}+n_{\beta}\bar{W}_{\alpha}+n_{\alpha}n_{\beta}\bar{Y},\label{eq:KKImC}
\end{equation}
with
\begin{eqnarray}
\bar{V}_{ij} & = & \left(N^{2}+\beta_{k}\beta^{k}\right)B_{ij}-\beta_{j}\beta^{k}B_{ik}-\beta^{k}\beta_{i}B_{kj}\nonumber \\
 &  & +h_{ij}\beta^{k}\beta^{l}B_{kl}-N\left(\epsilon_{kjs}\beta^{k}E_{i}^{\enskip s}+\epsilon_{kis}\beta^{k}E_{j}^{\enskip s}\right),\label{eq:Vbab}\\
\bar{W}_{i} & = & N\beta^{k}B_{ik}-\epsilon_{lis}\beta^{k}\beta^{l}E_{k}^{\enskip s},\label{eq:Wbb}\\
\bar{Y} & = & \beta^{i}\beta^{j}B_{ij}.\label{eq:Yb}
\end{eqnarray}
Rewriting (\ref{eq:Imsig}) we have also
\begin{equation}
\textrm{Im}\left(\sigma_{\nu}\right)=\bar{Z}_{\nu}+n_{\nu}\bar{T},\label{eq:ImS}
\end{equation}
with
\begin{eqnarray}
\bar{Z}_{i} & = & -\epsilon_{ijk}\left(2K_{l}^{\enskip k}\beta^{l}\beta^{j}-2\beta^{j}D^{k}N-ND^{j}\beta^{k}\right),\label{eq:Zbv}\\
\bar{T} & = & \epsilon_{ijk}\beta^{i}D^{j}\beta^{k},\label{eq:Tb}
\end{eqnarray}
so
\begin{eqnarray}
II_{\alpha\beta\nu} & = & 2\bar{V}_{\alpha\left[\beta\right.}\bar{Z}_{\left.\nu\right]}+2\bar{T}\bar{V}_{\alpha\left[\beta\right.}n_{\left.\nu\right]}\nonumber \\
 &  & +2n_{\alpha}\bar{W}_{\left[\beta\right.}\bar{Z}_{\left.\nu\right]}+2\bar{W}_{\alpha}n_{\left[\beta\right.}\bar{Z}_{\left.\nu\right]}\nonumber \\
 &  & +2\bar{T}\bar{W}_{\left[\beta\right.}n_{\left.\nu\right]}n_{\alpha}+2\bar{Y}n_{\alpha}n_{\left[\beta\right.}\bar{Z}_{\left.\nu\right]}.\label{eq:ReFTB}
\end{eqnarray}
Finally, using (\ref{eq:ReFTA}) and (\ref{eq:ReFTB}), and the antisymmetry
in $\beta$ and $\nu$, (\ref{eq:FTRe}) can be written
\begin{eqnarray}
\textrm{Re}\left(FT\right) & = & 2\left[V_{\alpha\left[\beta\right.}Z_{\left.\nu\right]}-\bar{V}_{\alpha\left[\beta\right.}\bar{Z}_{\left.\nu\right]}\right.\nonumber \\
 &  & +\left(TV_{\alpha\left[\beta\right.}-\bar{T}\bar{V}_{\alpha\left[\beta\right.}-W_{\alpha}Z_{\left[\beta\right.}+\bar{W}_{\alpha}\bar{Z}_{\left[\beta\right.}\right)n_{\left.\nu\right]}\nonumber \\
 &  & +\left(TW_{\left[\beta\right.}-\bar{T}\bar{W}_{\left[\beta\right.}-YZ_{\left[\beta\right.}+\bar{Y}\bar{Z}_{\left[\beta\right.}\right)n_{\left.\nu\right]}n_{\alpha}\nonumber \\
 &  & +\left.n_{\alpha}\left(W_{\left[\beta\right.}Z_{\left.\nu\right]}-\bar{W}_{\left[\beta\right.}\bar{Z}_{\left.\nu\right]}\right)\right].\label{eq:ReFT2}
\end{eqnarray}
For the imaginary part we do the same :
\begin{eqnarray}
\textrm{Im}\left(FT\right) & = & \xi^{\mu}\xi^{\rho}\textrm{Re}\left(\mathcal{C}_{\mu\alpha\rho\left[\beta\right.}\right)\textrm{Im}\left(\sigma_{\left.\nu\right]}\right)\nonumber \\
 &  & +\xi^{\mu}\xi^{\rho}\textrm{Im}\left(\mathcal{C}_{\mu\alpha\rho\left[\beta\right.}\right)\textrm{Re}\left(\sigma_{\left.\nu\right]}\right),\label{eq:ImFT}
\end{eqnarray}
and we only have to use (\ref{eq:KKReC}) with (\ref{eq:ImS}), and
(\ref{eq:KKImC}) with (\ref{eq:ReS}) to obtain
\begin{eqnarray}
\textrm{Im}\left(FT\right) & = & 2\left[V_{\alpha\left[\beta\right.}\bar{Z}_{\left.\nu\right]}+\bar{V}_{\alpha\left[\beta\right.}Z_{\left.\nu\right]}\right.\nonumber \\
 &  & +\left(\bar{T}V_{\alpha\left[\beta\right.}+T\bar{V}_{\alpha\left[\beta\right.}-W_{\alpha}\bar{Z}_{\left[\beta\right.}-\bar{W}_{\alpha}Z_{\left[\beta\right.}\right)n_{\left.\nu\right]}\nonumber \\
 &  & +\left(\bar{T}W_{\left[\beta\right.}+T\bar{W}_{\left[\beta\right.}-Y\bar{Z}_{\left[\beta\right.}-\bar{Y}Z_{\left[\beta\right.}\right)n_{\left.\nu\right]}n_{\alpha}\nonumber \\
 &  & +\left.n_{\alpha}\left(W_{\left[\beta\right.}\bar{Z}_{\left.\nu\right]}+\bar{W}_{\left[\beta\right.}Z_{\left.\nu\right]}\right)\right],\label{eq:ImFT2}
\end{eqnarray}
where $V_{\alpha\beta}$, $\bar{V}_{\alpha\beta}$, $Z_{\nu}$, $\bar{Z}_{\nu}$,
$T$, $\bar{T}$, $W_{\alpha}$, $\bar{W}_{\alpha}$, $Y$ and $\bar{Y}$
are given respectively by (\ref{eq:Vab}), (\ref{eq:Vbab}), (\ref{eq:Zv}),
(\ref{eq:Zbv}), (\ref{eq:T}), (\ref{eq:Tb}), (\ref{eq:Wb}), (\ref{eq:Wbb}),
(\ref{eq:Y}) and (\ref{eq:Yb}). 

\subsubsection{Second term}

We obtain the same kind of decomposition for the real part and the
imaginary part of the second term
\begin{eqnarray}
\textrm{Re}\left(ST\right) & = & \gamma_{\alpha\beta}\underset{I'_{\alpha\beta\nu}}{\underbrace{\xi^{\mu}\textrm{Re}\left(\mathcal{C}_{\nu\mu\rho\delta}\right)\textrm{Re}\left(\mathcal{F}^{\rho\delta}\right)}}\nonumber \\
 &  & -\gamma_{\alpha\beta}\underset{II'_{\alpha\beta\nu}}{\underbrace{\xi^{\mu}\textrm{Im}\left(\mathcal{C}_{\nu\mu\rho\delta}\right)\textrm{Im}\left(\mathcal{F}^{\rho\delta}\right)}},\label{eq:ReST-1}
\end{eqnarray}
but the contractions are a little more difficult to do for this term.
Using (\ref{eq:KV}), (\ref{eq:Wt2}) and (\ref{eq:ReF}) to calculate
$I'_{\alpha\beta\nu}$ and using (\ref{eq:KV}), (\ref{eq:ImC}),
(\ref{eq:ImF}) and the properties of the spatial volume form : (\ref{eq:EpsEps1})
and (\ref{eq:EpsEps2}) for $II'_{\alpha\beta\nu}$, we obtain 
\begin{equation}
\textrm{Re}\left(\gamma_{\alpha\beta}\xi^{\mu}\mathcal{C}_{\nu\mu\rho\delta}\mathcal{F}^{\rho\delta}\right)=\gamma_{\alpha\beta}\left(L_{\nu}+n_{\nu}M\right),\label{eq:ReST}
\end{equation}
with
\begin{eqnarray}
L_{i} & = & 2\left(D^{j}\beta^{k}-D^{k}\beta^{j}\right)\beta_{k}E_{ij}+2\beta^{k}E_{jk}\left(D_{i}\beta^{j}-D^{j}\beta_{i}\right)\nonumber \\
 &  & -2N\epsilon_{jkl}B_{i}^{\enskip l}D^{j}\beta^{k}+2A_{il}\left(\beta_{m}K^{lm}-D^{l}N\right),\label{eq:Lv}\\
M & = & -2\epsilon_{ijk}\beta^{l}B_{l}^{\enskip k}D^{i}\beta^{j}-4\beta^{i}E_{ij}\left(K^{jk}\beta_{k}-D^{j}N\right),\label{eq:M}
\end{eqnarray}
where 
\begin{equation}
A_{il}=\epsilon_{ikj}\beta^{k}B_{\enskip l}^{j}-\epsilon_{ijl}\beta_{k}B^{jk}-\epsilon_{jkl}\beta^{k}B_{\enskip i}^{j}-2NE_{il}.\label{eq:Ail}
\end{equation}
We can also rewrite (\ref{eq:gamma31}) in the following form
\begin{equation}
\gamma_{\alpha\beta}=G_{\alpha\beta}+N\left(\beta_{\beta}n_{\alpha}+\beta_{\alpha}n_{\beta}\right)+n_{\alpha}n_{\beta}\beta_{i}\beta^{i},\label{eq:gamma2}
\end{equation}
with
\begin{equation}
G_{ij}=\left(N^{2}-\beta_{k}\beta^{k}\right)h_{ij}+\beta_{i}\beta_{j},\label{eq:Fab2}
\end{equation}
so we have
\begin{eqnarray}
\textrm{Re}\left(ST\right) & = & 2G_{\alpha\left[\beta\right.}L_{\left.\nu\right]}+2Nn_{\alpha}\beta_{\left[\beta\right.}L_{\left.\nu\right]}\nonumber \\
 &  & +2\left(MG_{\alpha\left[\beta\right.}-N\beta_{\alpha}L_{\left[\beta\right.}\right)n_{\left.\nu\right]}\nonumber \\
 &  & +2n_{\alpha}\left(MN\beta_{\left[\beta\right.}-\beta_{i}\beta^{i}L_{\left[\beta\right.}\right)n_{\left.\nu\right]}.\label{eq:ReST3}
\end{eqnarray}
We do the same for the imaginary part 
\begin{eqnarray}
\textrm{Im}\left(ST\right) & = & \gamma_{\alpha\beta}\underset{III'_{\alpha\beta\nu}}{\underbrace{\xi^{\mu}\textrm{Re}\left(\mathcal{C}_{\nu\mu\rho\delta}\right)\textrm{Im}\left(\mathcal{F}^{\rho\delta}\right)}}\nonumber \\
 &  & +\gamma_{\alpha\beta}\underset{IV'_{\alpha\beta\nu}}{\underbrace{\xi^{\mu}\textrm{Im}\left(\mathcal{C}_{\nu\mu\rho\delta}\right)\textrm{Re}\left(\mathcal{F}^{\rho\delta}\right)}}.\label{eq:ImST-1}
\end{eqnarray}
To calculate $IV'_{\alpha\beta\nu}$ we just have to take the formula
of $I'_{\alpha\beta\nu}$ and make the changes $E\rightarrow B$ and
$B\rightarrow-E$, and to calculate $III'_{\alpha\beta\nu}$ we take
$II'_{\alpha\beta\nu}$ and make the changes $B\rightarrow E$ and
$E\rightarrow-B$, so we obtain
\begin{eqnarray}
\textrm{Im}\left(ST\right) & = & 2G_{\alpha\left[\beta\right.}\bar{L}_{\left.\nu\right]}+2Nn_{\alpha}\beta_{\left[\beta\right.}\bar{L}_{\left.\nu\right]}\nonumber \\
 &  & +2\left(\bar{M}G_{\alpha\left[\beta\right.}-N\beta_{\alpha}\bar{L}_{\left[\beta\right.}\right)n_{\left.\nu\right]}\nonumber \\
 &  & +2n_{\alpha}\left(\bar{M}N\beta_{\left[\beta\right.}-\beta_{i}\beta^{i}\bar{L}_{\left[\beta\right.}\right)n_{\left.\nu\right]},\label{eq:ImST}
\end{eqnarray}
with
\begin{eqnarray}
\bar{L}_{i} & = & 2\left(D^{j}\beta^{k}-D^{k}\beta^{j}\right)\beta_{k}B_{ji}+2\left(D_{i}\beta^{j}-D^{j}\beta_{i}\right)\beta^{k}B_{jk}\nonumber \\
 &  & +2N\epsilon_{jkl}E_{i}^{\enskip l}D^{j}\beta^{k}+2\bar{A}_{il}\left(\beta_{m}K^{lm}-D^{l}N\right),\label{eq:Lbv}\\
\bar{M} & = & 2\epsilon_{ijk}\beta^{l}E_{l}^{\enskip k}\left(D^{i}\beta^{j}\right)-4\beta^{i}B_{ij}\left(K^{jk}\beta_{k}-D^{j}N\right),\label{eq:Mb}
\end{eqnarray}
where 
\begin{equation}
\bar{A}_{il}=\epsilon_{ijl}\beta_{k}E^{jk}+\epsilon_{jkl}\beta^{k}E_{\enskip i}^{j}-\epsilon_{ijk}\beta^{j}E_{\enskip l}^{k}-2NB_{li}.\label{eq:Abil}
\end{equation}

\subsubsection{Final decomposition}

Gathering all those calculations, we obtain the real part of the decomposition
of the Simon-Mars tensor coming from (\ref{eq:ReFT2}) (with a factor
4) and (\ref{eq:ReST3}) :
\begin{eqnarray}
\textrm{Re}\left(S_{\alpha\left[\beta\nu\right]}\right) & = & 2\left(S_{\alpha\beta\nu}^{1}+S_{\alpha\beta}^{2}n_{\nu}-S_{\alpha\nu}^{2}n_{\beta}\right.\label{eq:ReSF}\\
 &  & \left.+S_{\beta\nu}^{3}n_{\alpha}+S_{\beta}^{4}n_{\alpha}n_{\nu}-S_{\nu}^{4}n_{\alpha}n_{\beta}\right),\nonumber 
\end{eqnarray}
with
\begin{eqnarray}
S_{ijk}^{1} & = & 4\left(V_{i\left[j\right.}Z_{\left.k\right]}-\bar{V}_{i\left[j\right.}\bar{Z}_{\left.k\right]}\right)+G_{i\left[j\right.}L_{\left.k\right]},\label{eq:R1}\\
S_{ij}^{2} & = & 4\left(TV_{ij}-\bar{T}\bar{V}_{ij}+\bar{W}_{i}\bar{Z}_{j}-W_{i}Z_{j}\right)\nonumber \\
 &  & +MG_{ij}-N\beta_{i}L_{j},\label{eq:R2}\\
S_{ij}^{3} & = & 4\left(W_{\left[i\right.}Z_{\left.j\right]}-\bar{W}_{\left[i\right.}\bar{Z}_{\left.j\right]}\right)+N\beta_{\left[i\right.}L_{\left.j\right]},\label{eq:R3}\\
S_{i}^{4} & = & 4\left(TW_{i}-\bar{T}\bar{W}_{i}+\bar{Y}\bar{Z}_{i}-YZ_{i}\right)\nonumber \\
 &  & +N\beta_{i}M-\beta_{l}\beta^{l}L_{i}.\label{eq:R4}
\end{eqnarray}
We see that $S_{ijk}^{1}$ is antisymmetric in its 2 last indices
and that $S_{ij}^{3}$ is antisymmetric. \\

We do the same for the imaginary part coming from (\ref{eq:ImFT2})
(with also a factor 4) and (\ref{eq:ImST}) :
\begin{eqnarray}
\textrm{Im}\left(S_{\alpha\beta\nu}\right) & = & 2\left(\bar{S}_{\alpha\beta\nu}^{1}+\bar{S}_{\alpha\beta}^{2}n_{\nu}-\bar{S}_{\alpha\nu}^{2}n_{\beta}\right.\label{eq:ImSF}\\
 &  & \left.+\bar{S}_{\beta\nu}^{3}n_{\alpha}+\bar{S}_{\beta}^{4}n_{\alpha}n_{\nu}-\bar{S}_{\nu}^{4}n_{\alpha}n_{\beta}\right),\nonumber 
\end{eqnarray}
with
\begin{eqnarray}
\bar{S}_{ijk}^{1} & = & 4\left(V_{i\left[j\right.}\bar{Z}_{\left.k\right]}+\bar{V}_{i\left[j\right.}Z_{\left.k\right]}\right)+G_{i\left[j\right.}\bar{L}_{\left.k\right]},\label{eq:Rb1}\\
\bar{S}_{ij}^{2} & = & 4\left(\bar{T}V_{ij}+T\bar{V}_{ij}-W_{i}\bar{Z}_{j}-\bar{W}_{i}Z_{j}\right)\nonumber \\
 &  & +G_{ij}\bar{M}-N\beta_{i}\bar{L}_{j},\label{eq:Rb2}\\
\bar{S}_{ij}^{3} & = & 4\left(W_{\left[i\right.}\bar{Z}_{\left.j\right]}+\bar{W}_{\left[i\right.}Z_{\left.j\right]}\right)+N\beta_{\left[i\right.}\bar{L}_{\left.j\right]},\label{eq:Rb3}\\
\bar{S}_{i}^{4} & = & 4\left(\bar{T}W_{i}+T\bar{W}_{i}-Y\bar{Z}_{i}-\bar{Y}Z_{i}\right)\nonumber \\
 &  & +N\beta_{i}\bar{M}-\beta_{l}\beta^{l}\bar{L}_{i}.\label{eq:Rb4}
\end{eqnarray}
All these terms must be zero for Kerr spacetime (this is checked in
\ref{sub:Kerr-black-hole}), the goal is to compute them for other
spacetimes. But, before doing so, we build two scalar fields to be
able to compare coordinate independent quantities, according to the
spirit of general relativity. 

\subsection{Simon-Mars scalars}

The simplest scalar we can form with the Simon-Mars tensor is its
``square'' : 
\begin{equation}
S_{\alpha\beta\nu}S^{\alpha\beta\nu}.\label{eq:Square}
\end{equation}
We decompose it in two scalars, the absolute value of its real part
and of its imaginary part, within the 3+1 formalism using the decomposition
of the Simon-Mars tensor given in the preceding section. \\

First let us consider the real part using (\ref{eq:ReSF}) and (\ref{eq:nn})
and the symmetries. All the spatial indices give zero contracted with
$\vec{n}$, so we obtain
\begin{eqnarray}
\ss & = & \left|\textrm{Re}\left(S_{\alpha\beta\nu}S^{\alpha\beta\nu}\right)\right|\nonumber \\
 & = & \left|4\left[S_{ijk}^{1}S^{1\: ijk}-\bar{S}_{ijk}^{1}\bar{S}^{1\: ijk}-2\left(S_{ij}^{2}S^{2\: ij}-\bar{S}_{ij}^{2}\bar{S}^{2\: ij}\right)\right.\right.\nonumber \\
 &  & \left.\left.-S_{ij}^{3}S^{3\: ij}+\bar{S}_{ij}^{3}\bar{S}^{3\: ij}+2\left(S_{i}^{4}S^{4\: i}-\bar{S}_{i}^{4}\bar{S}^{4\: i}\right)\right]\right|,\label{eq:1/4ReS}
\end{eqnarray}
and doing the same for the imaginary part, we obtain 
\begin{eqnarray}
\bar{\ss} & = & \left|\textrm{Im}\left(S_{\alpha\beta\nu}S^{\alpha\beta\nu}\right)\right|\nonumber \\
 & = & \left|4\left[S_{ijk}^{1}\bar{S}^{1\: ijk}+\bar{S}_{ijk}^{1}S^{1\: ijk}-2\left(S_{ij}^{2}\bar{S}^{2\: ij}+\bar{S}_{ij}^{2}S^{2\: ij}\right)\right.\right.\nonumber \\
 &  & \left.\left.-S_{ij}^{3}\bar{S}^{3\: ij}-\bar{S}_{ij}^{3}S^{3\: ij}+2\left(S_{i}^{4}\bar{S}^{4\: i}+\bar{S}_{i}^{4}S^{4\: i}\right)\right]\right|.\label{eq:1/4ImS}
\end{eqnarray}
We will also calculate these scalars for different spacetimes.

\section{\label{sec:Axisymmetric-spacetimes}Axisymmetric spacetimes}

Generally, alternatives to the Kerr Black Hole spacetime are axisymmetric,
so in this part we consider the specific case of stationary and axisymmetric
spacetimes. We use the quasi-isotropic coordinate system which is
adapted to these symmetries. 

\subsection{Quasi-isotropic coordinates\label{sub:Quasi-isotropic-coordinates}}

This work deals with stationary, axisymmetric and circular spacetimes.
For spacetimes with those symmetries, we can use quasi-isotropic coordinates
(see \cite{Rotstar,Wald}). In these coordinates, the line
element can be written
\begin{eqnarray}
\textrm{d}s^{2} & = & -N^{2}\textrm{d}t^{2}+A\left(\textrm{d}r^{2}+r^{2}\textrm{d}\theta^{2}\right)\nonumber \\
 &  & +B^{2}r^{2}\sin^{2}\theta\left(\textrm{d}\varphi+\beta^{\varphi}\textrm{d}t\right)^{2}.\label{eq:metQI}
\end{eqnarray}
Comparing with (\ref{eq:met}), we identify the 3+1 quantities of
this metric : $N$ is the lapse, the shift is given by $\beta^{i}=\left(0,0,\beta^{\varphi}\right)$
and the spatial metric reads 
\begin{equation}
h_{ij}=\left(\begin{array}{ccc}
A^{2} & 0 & 0\\
0 & A^{2}r^{2} & 0\\
0 & 0 & B^{2}r^{2}\sin^{2}\theta
\end{array}\right),\label{eq:ssm}
\end{equation}
where $\beta^{\varphi}$, $A$ and $B$ depend only on $r$ and $\theta$.\\

In these coordinates, we can easily check that we have always
\begin{equation}
T=\bar{T}=0.\label{eq:T0}
\end{equation}
Furthermore, we have $R_{r\varphi}=R_{\theta\varphi}=0$ which implies
\begin{equation}
E_{r\varphi}=E_{\theta\varphi}=B_{r\varphi}=B_{\theta\varphi}=0,\label{eq:EH0}
\end{equation}
for Kerr (for which the stress energy tensor $T_{\mu\nu}=0$) and
for boson stars (for which $T_{r\varphi}=T_{\theta\varphi}=T_{rt}=T_{\theta t}=0$).
So in this case we have also always 
\begin{equation}
M=\bar{M}=0,\label{eq:M0}
\end{equation}
and only the following projections are non zero (the 2-tensors are
all symmetric) : $V_{rr}$, $V_{\theta\theta}$, $V_{\varphi\varphi}$,
$V_{r\theta}$, $\bar{V}_{rr}$, $\bar{V}_{\theta\theta}$, $\bar{V}_{\varphi\varphi}$,
$\bar{V}_{r\theta}$, $W_{\varphi}$, $\bar{W_{\varphi}}$, $Z_{r}$,
$Z_{\theta}$, $\bar{Z}_{r}$, $\bar{Z_{\theta}}$, $G_{rr}$, $G_{\theta\theta}$,
$G_{\varphi\varphi}$, $L_{r}$, $L_{\theta}$, $\bar{L}_{r}$, $\bar{L_{\theta}}$.
This implies the following equalities 
\begin{eqnarray}
S_{ij}^{2} & = & -2S_{ij}^{3},\label{eq:R23}\\
\bar{S}_{ij}^{2} & = & -2\bar{S}_{ij}^{3}.\label{eq:iR23}
\end{eqnarray}
Thus, we need only to compute the tensors $S_{ijk}^{1}$, $S_{ij}^{2}$,
and $S_{i}^{4}$ given by (\ref{eq:R1}), (\ref{eq:R2}) and (\ref{eq:R4})
for the real part and $\bar{S}_{ijk}^{1}$, $\bar{S}_{ij}^{2}$, and
$\bar{S}_{i}^{4}$ given by (\ref{eq:Rb1}), (\ref{eq:Rb2}) and (\ref{eq:Rb4})
for the imaginary part. Furthermore, the non zero components of these
tensors are : $S_{rr\theta}^{1}$, $S_{\theta r\theta}^{1}$, $S_{\varphi r\varphi}^{1}$,
$S_{\varphi\theta\varphi}^{1}$, $S_{\varphi r}^{2}$, $S_{\varphi\theta}^{2}$,
$S_{r}^{4}$, $S_{\theta}^{4}$ and exactly the same for the imaginary
part.\\

Finally, because of (\ref{eq:R23}) and (\ref{eq:iR23}), we can write
$\ss$ and $\bar{\ss}$ as 
\begin{eqnarray}
\ss & = & \left|4\left[S_{ijk}^{1}S^{1\: ijk}-\bar{S}_{ijk}^{1}\bar{S}^{1\: ijk}+2\left(S_{i}^{4}S^{4\: i}-\bar{S}_{i}^{4}\bar{S}^{4\: i}\right)\right]\right.\nonumber \\
 &  & \left.-9\left(S_{ij}^{2}S^{2\: ij}-\bar{S}_{ij}^{2}\bar{S}^{2\: ij}\right)\right|,\label{eq:defS1}\\
\bar{\ss} & = & \left|4\left[S_{ijk}^{1}\bar{S}^{1\: ijk}+\bar{S}_{ijk}^{1}S^{1\: ijk}+2\right]\left(S_{i}^{4}\bar{S}^{4\: i}+\bar{S}_{i}^{4}S^{4\: i}\right)\right.\nonumber \\
 &  & \left.-9\left(S_{ij}^{2}\bar{S}^{2\: ij}-\bar{S}_{ij}^{2}S^{2\: ij}\right)\right|.\label{eq:defS2}
\end{eqnarray}
Let us do the calculation for the simplest case : the Schwarzschild
spacetime.

\subsection{Schwarzschild spacetime}

For the spherically symmetric Schwarzschild spacetime, we have $K_{ij}=0$
and $\beta_{k}=0$. Thanks to this symmetry, the quasi-isotropic coordinates
become isotropic because $A=B$ in (\ref{eq:ssm}) such that 
\begin{equation}
h_{ij}=A^{2}f_{ij},\label{eq:HS}
\end{equation}
where $f_{ij}$ is a flat metric ((\ref{eq:HS}) means that the Schwarzschild
metric is conformally flat). In these coordinates, the electric (\ref{eq:E})
and magnetic (\ref{eq:B}) parts of the self-dual Weyl tensor are
given by 
\begin{eqnarray}
E_{ij} & = & R_{ij},\label{eq:EBS}\\
B_{ij} & = & 0.\label{eq:EBS2}
\end{eqnarray}
The only non null terms are (\ref{eq:Vab}), (\ref{eq:Zv}), (\ref{eq:Fab2})
and (\ref{eq:Lv})
\begin{eqnarray}
V_{ij} & = & N^{2}R_{ij},\label{eq:VS}\\
Z_{i} & = & 2ND_{i}N,\label{eq:ZS}\\
G_{ij} & = & N^{2}h_{ij},\label{eq:FS}\\
L_{i} & = & 4NR_{ij}D^{j}N,\label{eq:LS}
\end{eqnarray}
so in this case $S_{ijk}^{1}$ (\ref{eq:R1}) is the only component
non trivially zero :
\begin{eqnarray}
S_{ijk}^{1\:\textrm{Sch}} & = & 4V_{i\left[j\right.}Z_{\left.k\right]}+G_{i\left[j\right.}L_{\left.k\right]}\nonumber \\
 & = & 4N^{3}\left(2R_{i\left[j\right.}D_{\left.k\right]}N+h_{i\left[j\right.}R_{\left.k\right]l}D^{l}N\right).\label{eq:R1S2}
\end{eqnarray}
But, in isotropic coordinates, with (\ref{eq:HS}), we can calculate
(repeated indices are not summed here) 
\begin{eqnarray}
R_{r\varphi} & = & R_{\theta\varphi}=R_{r\theta}=0,\label{eq:Rrp}\\
R_{\theta\theta} & = & -\frac{h_{\theta\theta}h^{rr}}{2}R_{rr}=-\frac{r^{2}}{2}R_{rr},\label{eq:Rtt}\\
R_{\varphi\varphi} & = & -\frac{h_{\varphi\varphi}h^{rr}}{2}R_{rr}=-\frac{r^{2}\sin^{2}\theta}{2}R_{rr}.\label{eq:Rpp}
\end{eqnarray}
This is logical because the Ricci scalar for Schwarzschild is zero
so we have 
\begin{equation}
0=R=\frac{1}{A^{2}}\left(R_{rr}+\frac{R_{\theta\theta}}{r^{2}}+\frac{R_{\varphi\varphi}}{r^{2}\sin^{2}\theta}\right).\label{eq:Rs}
\end{equation}
Writing (\ref{eq:R1S2}) in components and using (\ref{eq:Rrp})-(\ref{eq:Rpp}), we conclude that 
\begin{equation}
S_{ijk}^{1\:\textrm{Sch}}=0,\label{eq:R1SC}
\end{equation}
which is the result we expect.\\

\section{\label{sec:Numerical-applications}Applications}

Once we have all the components (\ref{eq:R1}) to (\ref{eq:Rb4})
and the scalars (\ref{eq:defS1}) and (\ref{eq:defS2}), we can evaluate
them for different spacetimes and develop a quantification of the
``non-Kerness'' of those spacetimes. Thanks to the 3+1 decomposition,
we can compute the Simon-Mars tensor components and scalars in every
kind of stationary spacetimes. To illustrate this, we considered two
examples of purely numerical spacetimes, two analytical spacetimes
and a modified Kerr metric which is a parametric deviation from
Kerr. Each example required an adapted tool : for numerical solutions
we used codes based on the Kadath library \cite{Grand10} available on line \cite{Kadath},
and for analytic solutions we used the 
SageManifolds extension of the open-source computer algebra system Sage
\cite{GourgBM14,SageManifolds}. 

\subsection{Units}

In this section we use geometrical units for which $c=G=1$. In those
units, the dimension of the Simon-Mars tensor is $1/L^{3}$, so the
dimension of the two scalar fields $\ss$ and $\bar{\ss}$ is $1/L^{6}$.
To manipulate dimensionless quantities, we chose to scale by $M$,
the ADM (Arnowitt, Deser and Misner) mass of each spacetime considered.
In geometrical units, a length has the same dimension as a mass, so
in this part every length is given in units of the ADM mass, and $\ss$
means $\ss\times M^{6}$, same thing for $\bar\ss$.

\subsection{Numerical solutions of the Einstein equations}

First we present the case of Kerr spacetime in quasi-isotropic coordinates
\cite{Gourg}, which is analytic, to test the validity of the calculation.
Then, we apply it on rotating boson star spacetimes and rotating neutron
stars which are solutions of the Einstein equation with a matter content
solved by making use of the Kadath library (see \cite{GSG14} for boson stars and \cite{Gourg01}
for neutron stars).

\subsubsection{\label{sub:Kerr-black-hole}Kerr black hole}

First we had to encode Kerr spacetime in the Kadath library. The 3D
spacetime $\Sigma$ is decomposed into several numerical domains.
The first one starts outside the event horizon, which is located at
$r=\frac{1}{2}\sqrt{M^{2}-a^{2}}$ in quasi-isotropic coordinates,
and the last one is compactified. In each domain the fields are described
by their development on chosen basis functions (typically Chebyshev
polynomials). For each domain we define the lapse, shift and 3D metric.
For Kerr spacetime we choose to write these 3+1 quantities in quasi-isotropic
coordinates, given in \cite{Gourg}. We verify that the vacuum 3+1
Einstein equations (Hamiltonian constraint, momentum constraint and
evolution equation, cf \cite{Gourg12}) are well verified. \\

\begin{figure}[!hbtp]
\includegraphics[width=8cm]{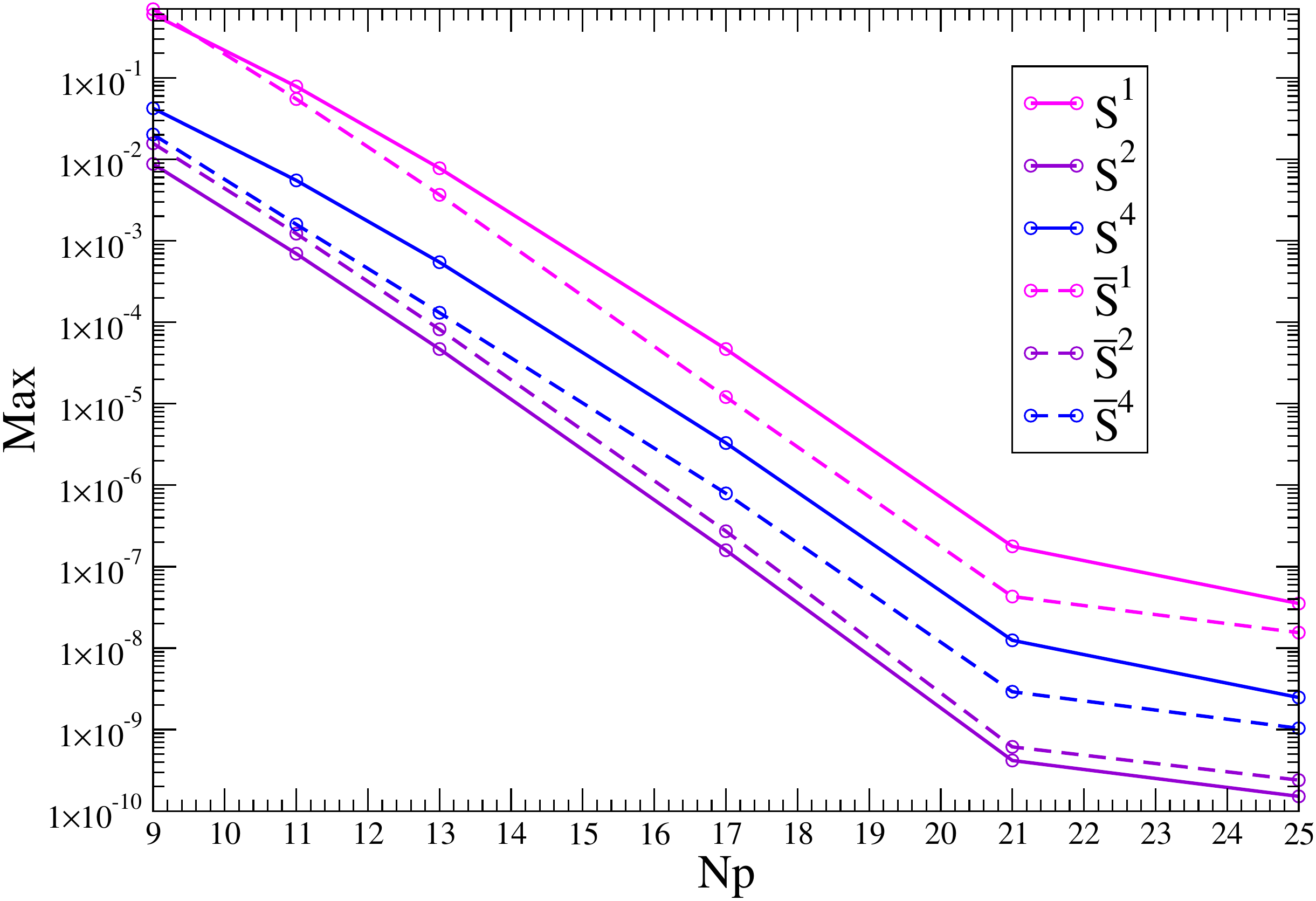}
\includegraphics[width=8cm]{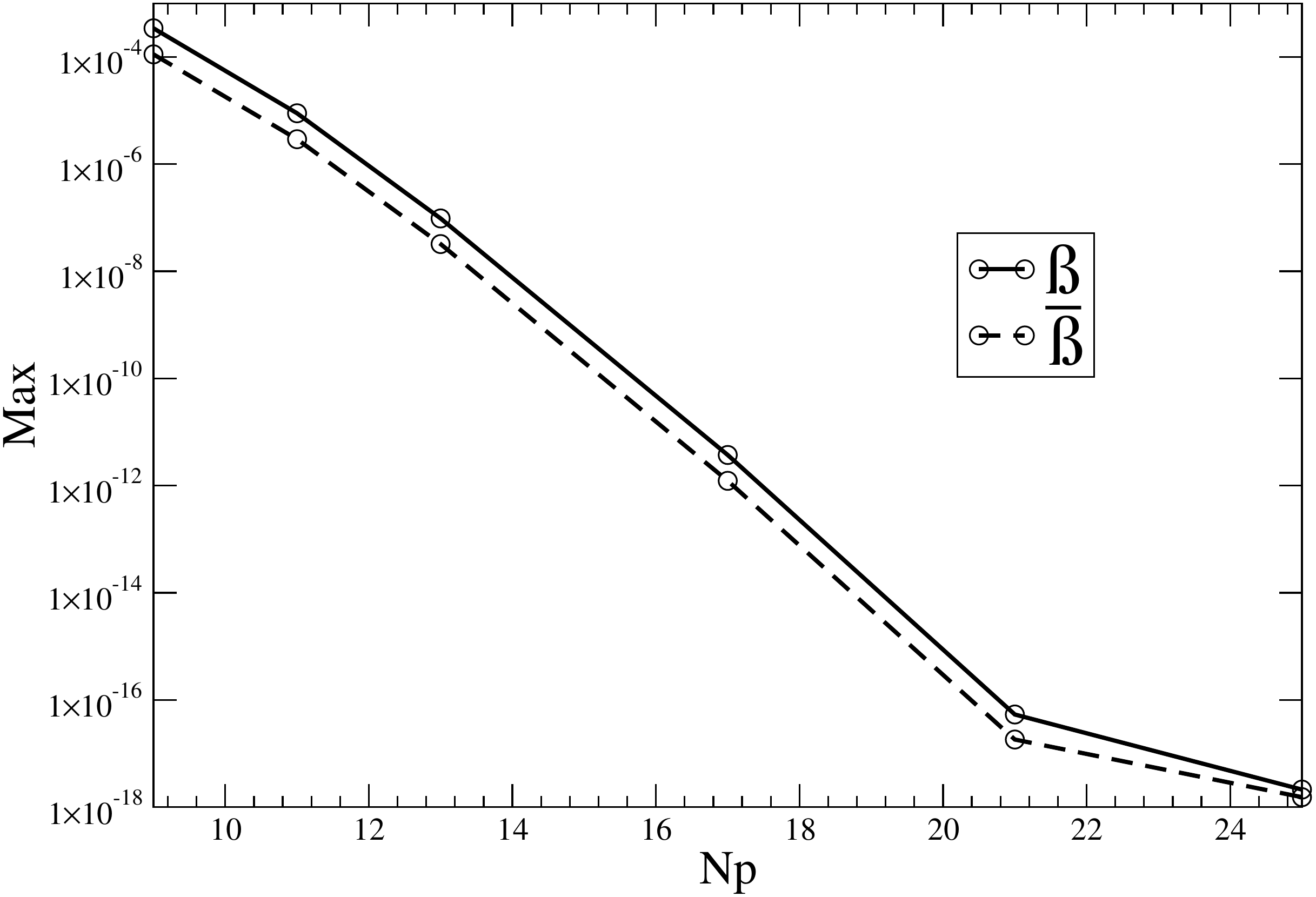}
\caption{Maximal value of each component of $S_{ijk}^{1}$, $S_{ij}^{2}$,
and $S_{i}^{4}$ and of $\bar{S}_{ijk}^{1}$, $\bar{S}_{ij}^{2}$,
and $\bar{S}_{i}^{4}$ in the first panel and of $\ss$ and $\bar{\ss}$ in
the second panel for the Kerr spacetime characterized by $M=1$ and
$a=0.8\: M$ as a function of the number of spectral coefficients
in both the radial and angular dimension (we chose the same number
for these two dimensions).\label{fig:SpasiKerr}}
\end{figure}

Then we want to verify that the Simon-Mars tensor is identically null
for Kerr spacetime, so we compute the 8 components (by virtue of (\ref{eq:R23})
and (\ref{eq:iR23}) only 6 of them are meaningful) of the
Simon-Mars tensor for Kerr spacetime with $a=0.8\: M$. In the Fig.~\ref{fig:SpasiKerr} is shown the maximal value of each component
of $S_{ijk}^{1}$, $S_{ij}^{2}$, $S_{i}^{4}$, $\bar{S}_{ijk}^{1}$, $\bar{S}_{ij}^{2}$,
and $\bar{S}_{i}^{4}$ and of the two
scalars $\ss$ and $\bar{\ss}$, in all domains (i.e. outside the
event horizon).

We can see that the maximal value of each component decreases exponentially
with the resolution, such a convergence is typical of spectral methods
and show that all those values are equal to zero. This successful
test validates our 3+1 decomposition of the Simon-Mars tensor, now
we consider other stationary, axisymmetric and asymptotically flat
spacetimes.

\subsubsection{Rotating boson stars}

Boson stars are localized configurations of a complex self-gravitating
field $\Phi$. Their study is motivated by the fact that they can play the
role of black hole mimickers \cite{Guzman}. For instance, this model
presents a viable alternative to the Kerr black hole for the description
of the Galactic Center. Mathematically, this is a solution to the
coupled system Einstein-Klein-Gordon equations 
\begin{eqnarray}
R_{\alpha\beta}-\frac{1}{2}Rg_{\alpha\beta} & = & 8\pi T_{\alpha\beta},\label{eq:Eeq}\\
\nabla_{\alpha}\nabla^{\alpha}\Phi & = & \frac{\textrm{d}V}{\textrm{d}\left|\Phi\right|^{2}}\Phi,\label{eq:KGeq}
\end{eqnarray}
where 
\begin{equation}
T_{\alpha\beta}=\nabla_{\left(\alpha\right.}\bar{\Phi}\nabla_{\left.\beta\right)}\Phi-\frac{1}{2}g_{\alpha\beta}\left[\nabla^{\mu}\bar{\Phi}\nabla_{\mu}\Phi+V\left(\left|\Phi\right|^{2}\right)\right].\label{eq:Tab}
\end{equation}
The potential can take different forms depending on the model we choose
for the boson star, here we consider only ``mini'' boson star formed
with a free field potential :

\begin{equation}
V\left(\left|\Phi\right|^{2}\right)=\frac{m^{2}}{\hbar^{2}}\left|\Phi\right|^{2},\label{eq:VBS1}
\end{equation}
where $m$ is the boson mass.\\

To solve the coupled equations (\ref{eq:Eeq}) and (\ref{eq:KGeq}),
the following ansatz is used :
\begin{equation}
\Phi=\phi\left(r,\theta\right)\exp\left[i\left(\omega t-k\varphi\right)\right].\label{eq:phiBS}
\end{equation}
A specific boson star is characterized then by the values of $\omega$,
which is a real parameter, and $k$, which is an integer (non-rotating
boson stars have $k=0$). Given this, the 3+1 decomposition of the
matter part (\ref{eq:rho})-(\ref{eq:Sab}) of
a boson star is 
\begin{eqnarray}
\rho & = & \left[\frac{\left(\omega+k\beta^{\varphi}\right)^{2}}{N^{2}}+k^{2}h^{\varphi\varphi}\right]\frac{\phi^{2}}{2}\nonumber \\
 &  & +\frac{h^{ij}}{2}\partial_{i}\phi\partial_{j}\phi+\frac{V}{2},\label{eq:rhoBS}\\
p_{\varphi} & = & \frac{k}{N}\left(\omega+k\beta^{\varphi}\right)\phi^{2},\label{eq:pBS}\\
S_{ij} & = & \partial_{\left(i\right.}\bar{\phi}\mathcal{\partial}_{\left.j\right)}\phi\nonumber \\
 &  & +\frac{\gamma_{ij}}{2}\left[\frac{\left(\omega+k\beta^{\varphi}\right)^{2}\phi^{2}}{N^{2}}-\partial_{l}\bar{\phi}\mathcal{\partial}^{l}\phi-V\right].\label{eq:SabBS}
\end{eqnarray}
We write then the Einstein-Klein-Gordon equations (\ref{eq:Eeq})
and (\ref{eq:KGeq}) in 3+1 form using quasi-isotropic coordinates
(see \cite{Rotstar}), and those equations are numerically solved
by Kadath (see \cite{GSG14} for details). More on boson stars can
be found in \cite{Rev, Rev 2} (and in references of \cite{GSG14}).\\

\begin{figure}[!hbtp]
\includegraphics[width=8cm]{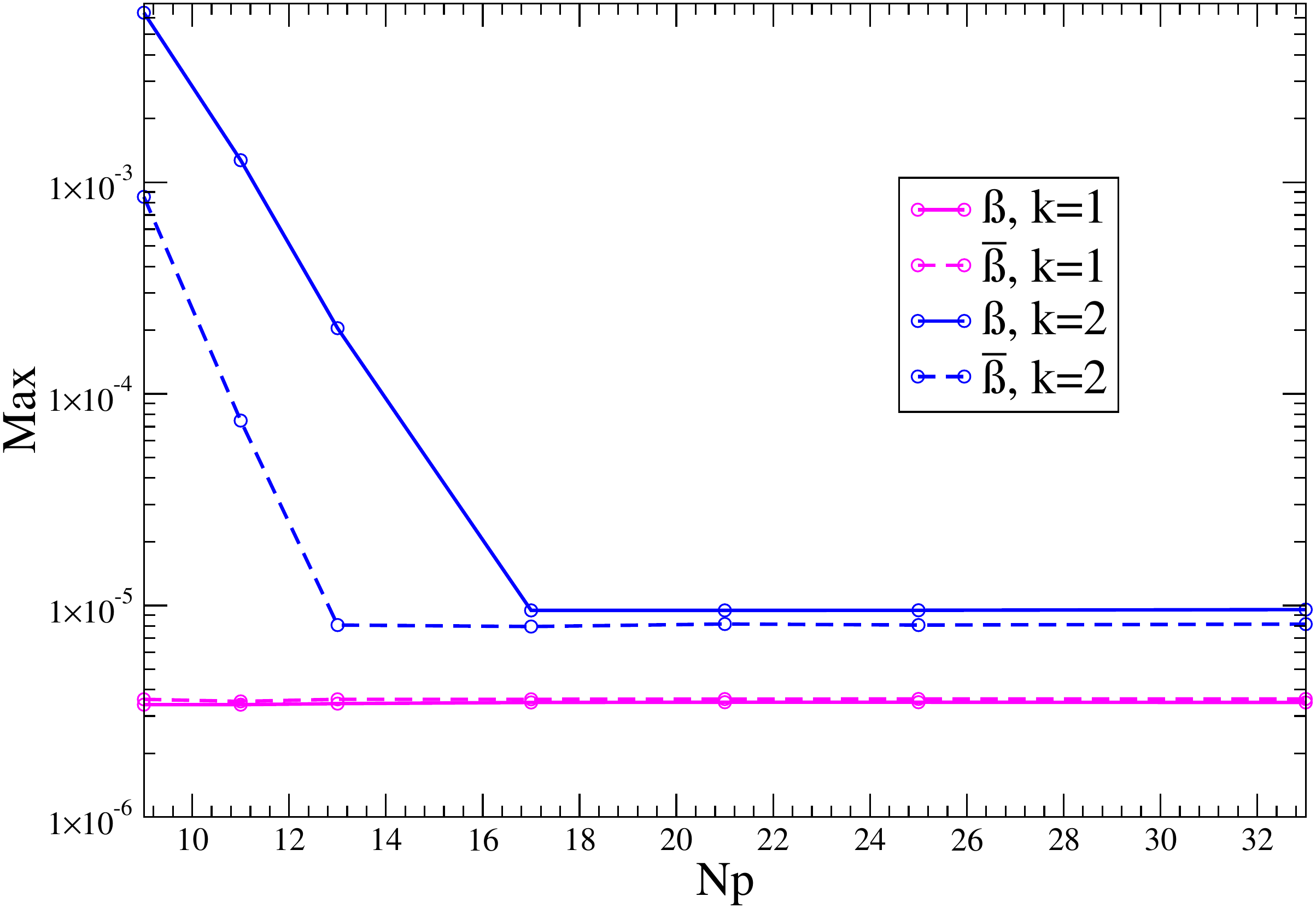}
\caption{Maximal values $\ss$ and $\bar{\ss}$ as functions of the number
of spectral coefficients in both the radial and angular dimension
for boson stars with the free field potential (\ref{eq:VBS1}) for
$\omega=0.8\, m/\hbar=1.05\, M^{-1}$ and $k=1,2$. \label{fig:SpasiBSConvk}}
\end{figure}

Let us plot in Fig.~\ref{fig:SpasiBSConvk} the same quantities
we did for Kerr in Fig.~\ref{fig:SpasiKerr}, that is to say the
maximal values of $\ss$ and $\bar{\ss}$ for several rotating boson
stars as functions of the resolution.\\

Contrary to the case of Kerr spacetime, the maximal values of $\ss$
and $\bar{\ss}$ do not depend on the resolution up to a certain precision.
The conclusion is that it is not zero. The rest of this paper, exploring
the same quantities for other spacetimes will permit us to give a
meaning to the maximal value found for $\ss$ and $\bar{\ss}$. Furthermore,
thanks to Fig.~\ref{fig:SpasiBSConvk}, we can choose a fine resolution
for the following plots, which will be $17$ points in $r$ and $\theta$.
\\

\begin{figure}[!hbtp]
\includegraphics[width=8cm]{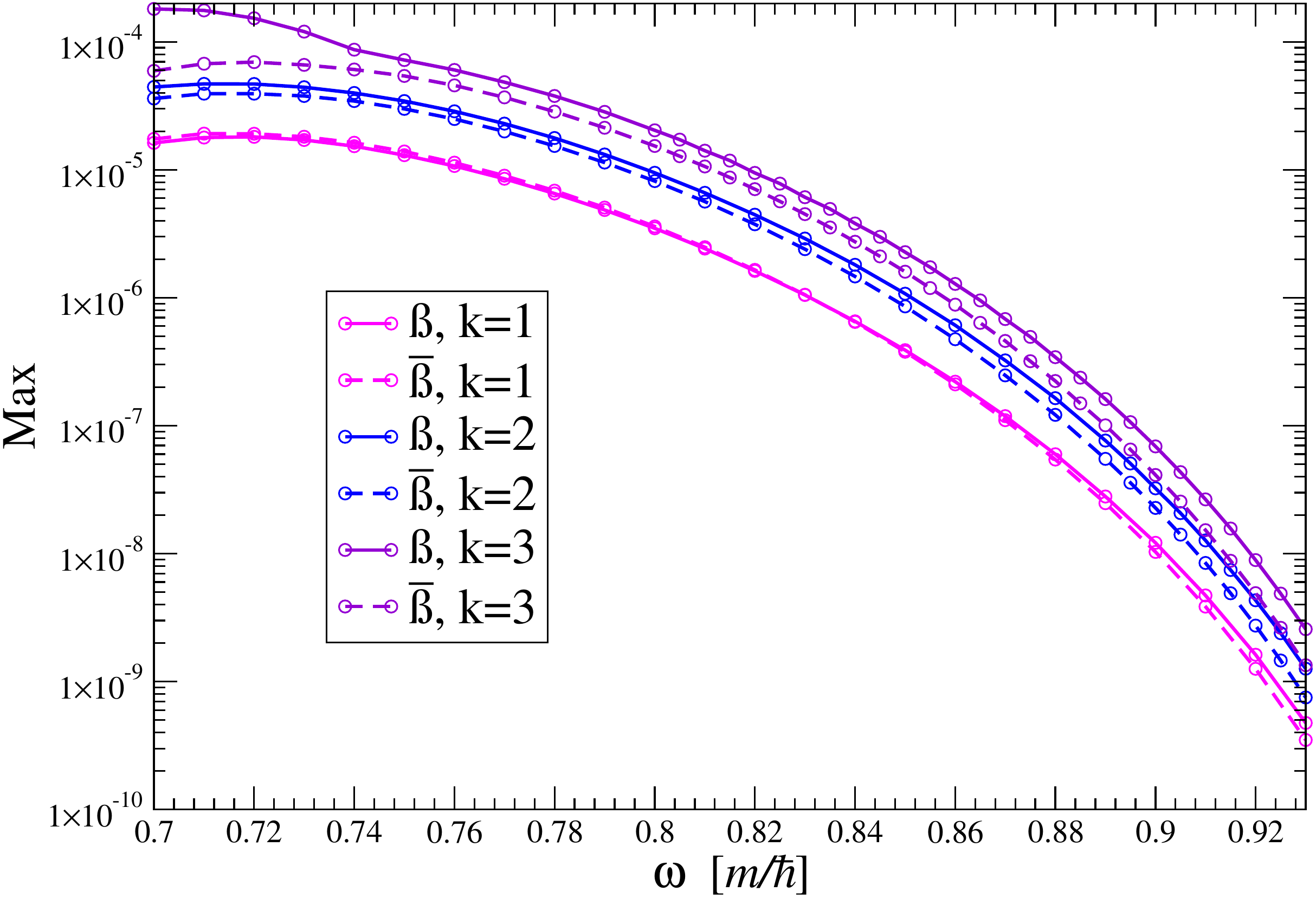}
\caption{Maximal values $\ss$ and $\bar{\ss}$ as functions of $\omega$ for
boson stars with $k=1,2,3$. As we recover Minkowski spacetime when $\omega\rightarrow1$
(in units of $\left[m/\hbar\right]$, see \cite{GSG14}), we expect
that these values tend to zero which is the case.\label{fig:SpasiBSomega}}
\end{figure}

We can also explore the global behavior of the maximal values of $\ss$
and $\bar{\ss}$ for different boson stars for a fixed resolution.
For instance, we plot in Fig.~\ref{fig:SpasiBSomega} those
values as functions of $\omega$ for various boson stars.\\

\begin{figure}[!hbtp]
\includegraphics[width=8cm]{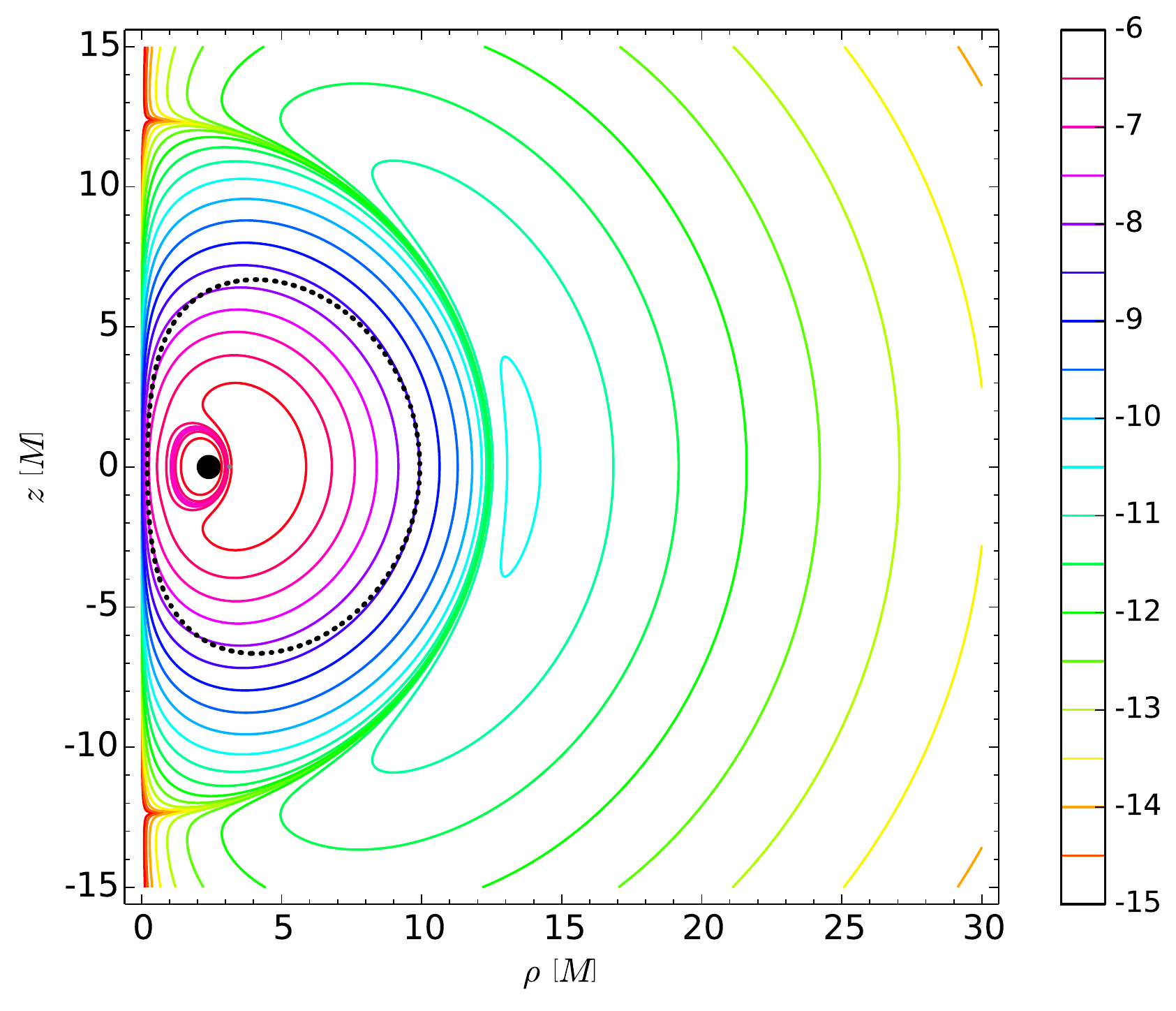}
\includegraphics[width=8cm]{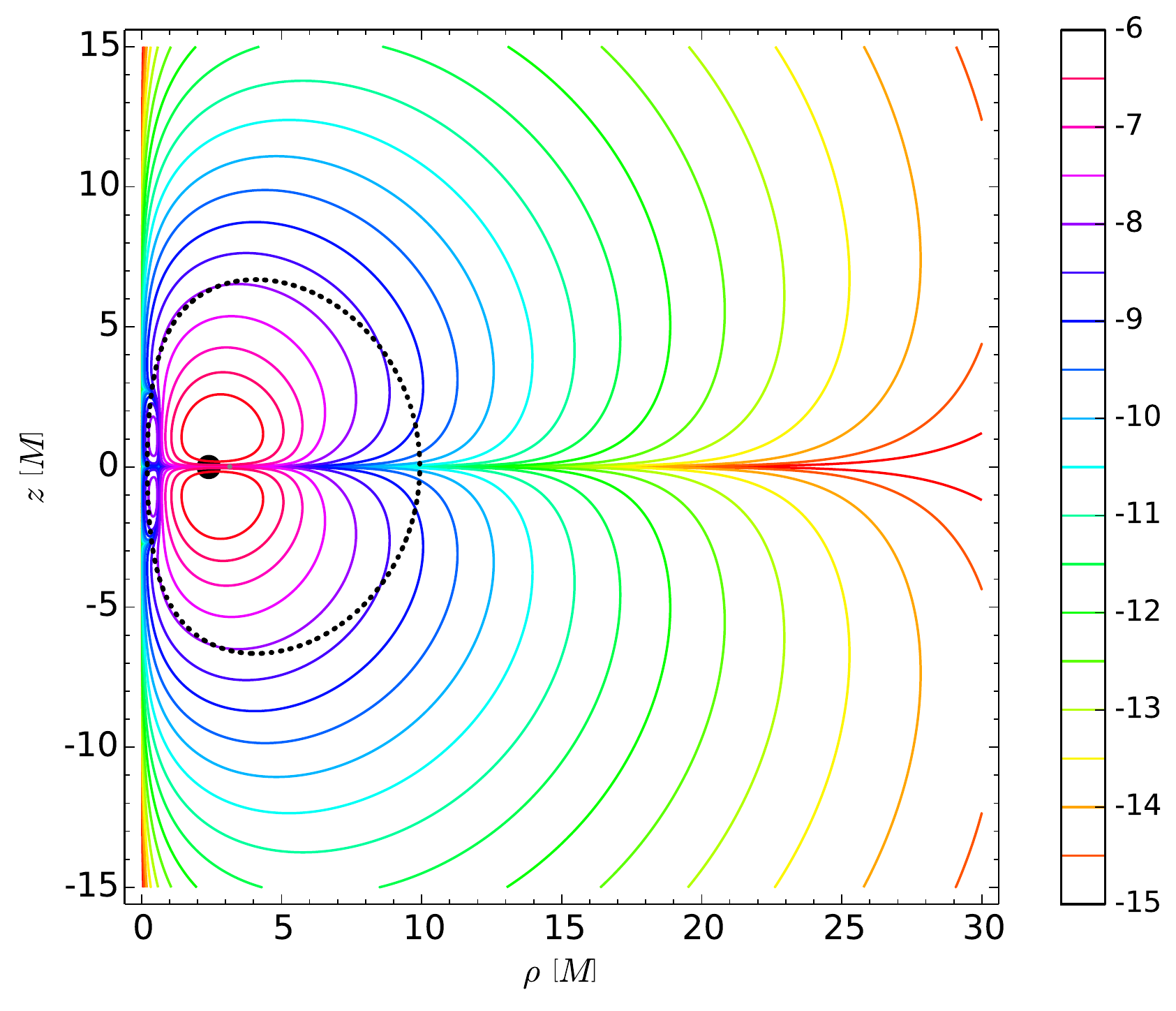}
\caption{Contour plot of $\log\left(\ss\right)$ and $\log\left(\bar{\ss}\right)$
as functions of $r$ and $z$ for a specific boson star characterized
by $\omega=1.05\, M^{-1}$ and $k=1$. The lengths are given in units
of the ADM mass of each boson star, to make the two plots comparable.
The black point marks the position of the maximal value of the boson
star field, and the dotted line indicate the positions where the field
take the value $\phi_{\textrm{max}}/10$. \label{fig:SpasiBSR12-1}}
\end{figure}

\begin{figure}[!hbtp]
\includegraphics[width=8cm]{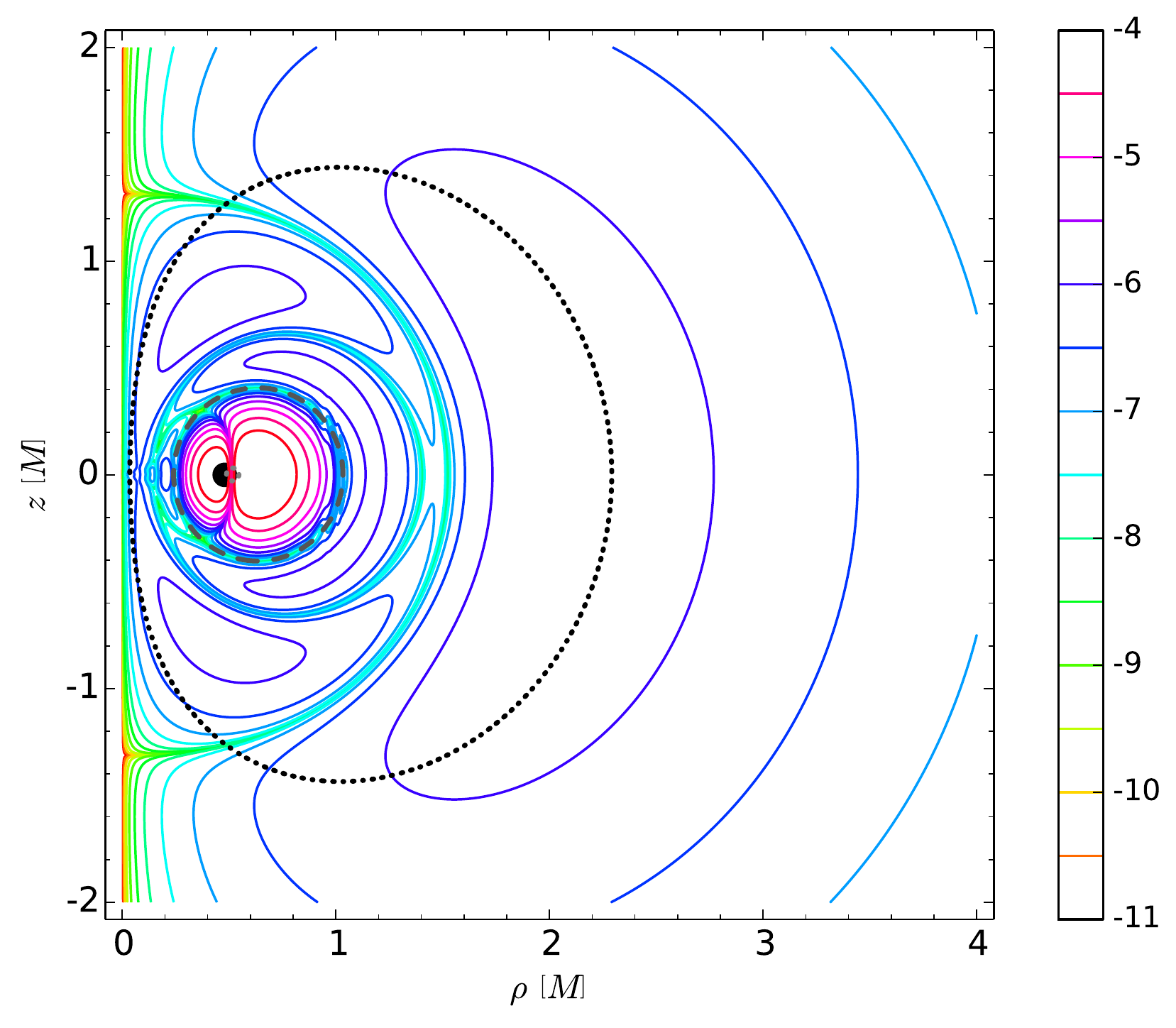}
\includegraphics[width=8cm]{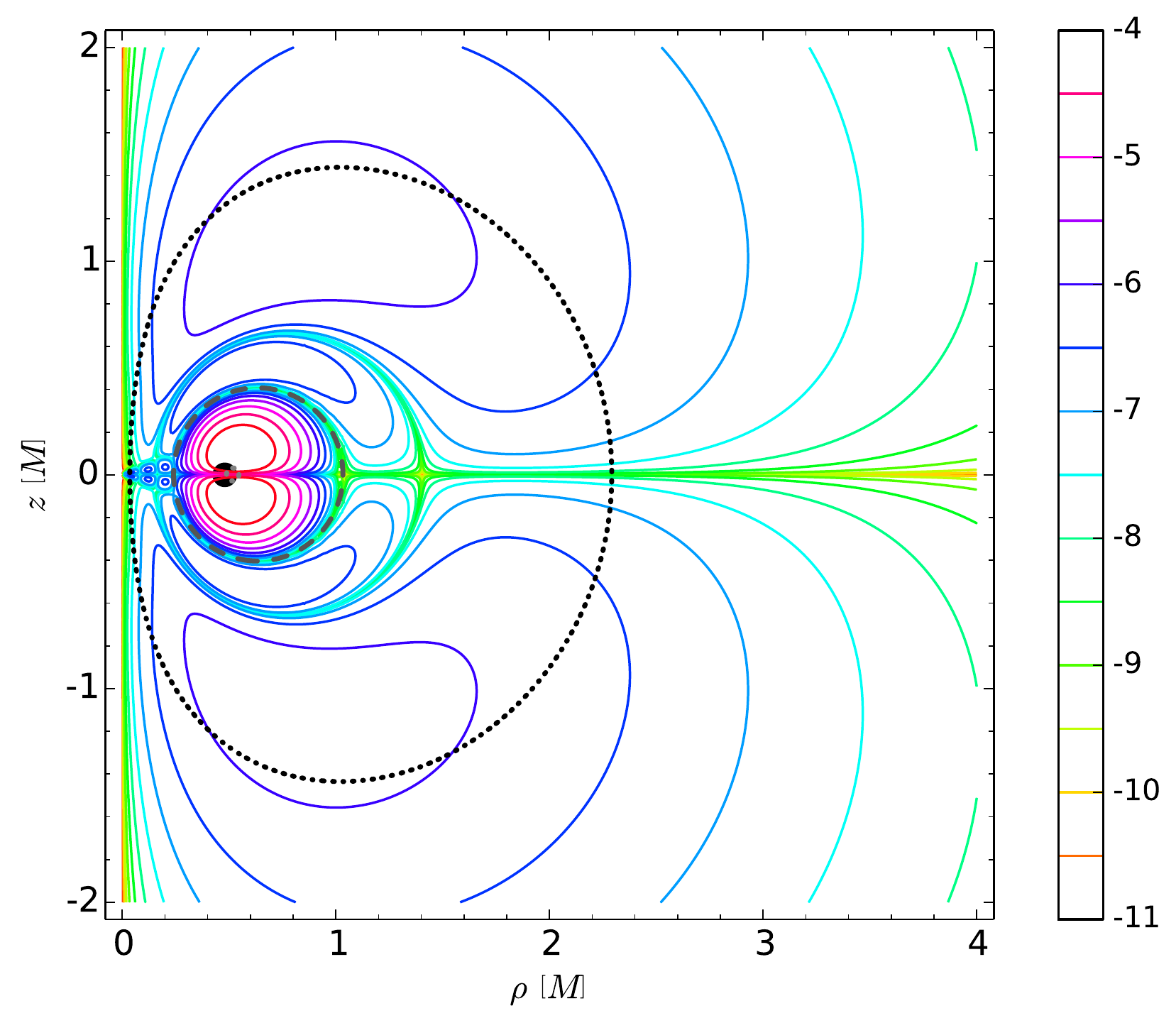}
\caption{Same figure as Fig.~\ref{fig:SpasiBSR12-1} for the boson star characterized
by $\omega=0.67\, M^{-1}$ and $k=1$. The grey dashed line
marks the location of the ergoregion. \label{fig:SpasiBSR12}}
\end{figure}

To see how the Simon-Mars scalars behave locally, we present in Fig.~\ref{fig:SpasiBSR12-1} and \ref{fig:SpasiBSR12} contour plots of
$\log\left(\ss\right)$ and $\log\left(\bar{\ss}\right)$ in the equatorial
plane as functions of the quasi-isotropic version of the Weyl-Papapetrou
coordinates : $r$ and $z$. As the scalar field $\phi$ decreases
exponentially fast after reaching its maximum, it makes sense to 
plot $\log\left(\ss\right)$ and $\log\left(\bar{\ss}\right)$.\\

\subsubsection{Rotating neutron stars}

Another example of stationary, axisymmetric and asymptotically flat
spacetime with a matter content, for which the metric can also be
expressed in quasi-isotropic coordinates is the model of rotating
neutron stars. Contrary to boson stars, the matter content is a perfect fluid. 
The stress
tensor has the following form 
\begin{equation}
T_{\alpha\beta}=\left(\varepsilon+p\right)u_{\alpha}u_{\beta}+pg_{\alpha\beta},\label{eq:Tpf}
\end{equation}
where $u^{\alpha}$ is the unit timelike vector field representing
the fluid 4-velocity, $\varepsilon$ and $p$ are the two scalar fields
representing respectively the energy density and the pressure of the
fluid. The 3+1 decomposition of the 4-velocity of the fluid with respect
to the Eulerian observer 4-velocity $\vec{n}$ is 
\begin{equation}
u^{\alpha}=\Gamma\left(n^{\alpha}+U^{\alpha}\right),\label{eq:upf}
\end{equation}
where $\Gamma=-n^{\mu}u_{\mu}$ is the Lorentz factor of the fluid
with respect to the Eulerian observer and the 3-velocity of the fluid
with respect also to this Eulerian observer is \cite{Rotstar}
\begin{equation}
U^{\alpha}=\frac{B}{N}\left(\Omega-N^{\varphi}\right)r\sin\theta,\label{eq:Upf}
\end{equation}
where $\Omega$ is the orbital angular velocity with respect to a
distant inertial observer. The normalization condition $u^{\mu}u_{\mu}=-1$
gives 
\begin{equation}
\Gamma=\frac{1}{\sqrt{1-U^{2}}}.\label{eq:gampf}
\end{equation}
So the 3+1 matter content (\ref{eq:rho})-(\ref{eq:Sab})
can be written 
\begin{eqnarray}
\rho & = & \Gamma^{2}\left(\varepsilon+p\right)-p\label{eq:rhoNS}\\
p_{\alpha} & = & \Gamma^{2}\left(\varepsilon+p\right)U_{\alpha}\label{eq:pNS}\\
S_{\alpha\beta} & = & \Gamma^{2}\left(\varepsilon+p\right)U_{\alpha}U_{\beta}+p\, h_{\alpha\beta}.\label{eq:SNS}
\end{eqnarray}

In order to close the system, we consider a polytropic equation of
state with $\gamma=2$. To read more about neutron stars and how these
objects are computed numerically see \cite{Gourg01} and references
therein.\\

\begin{figure}[!hbtp]
\includegraphics[width=8cm]{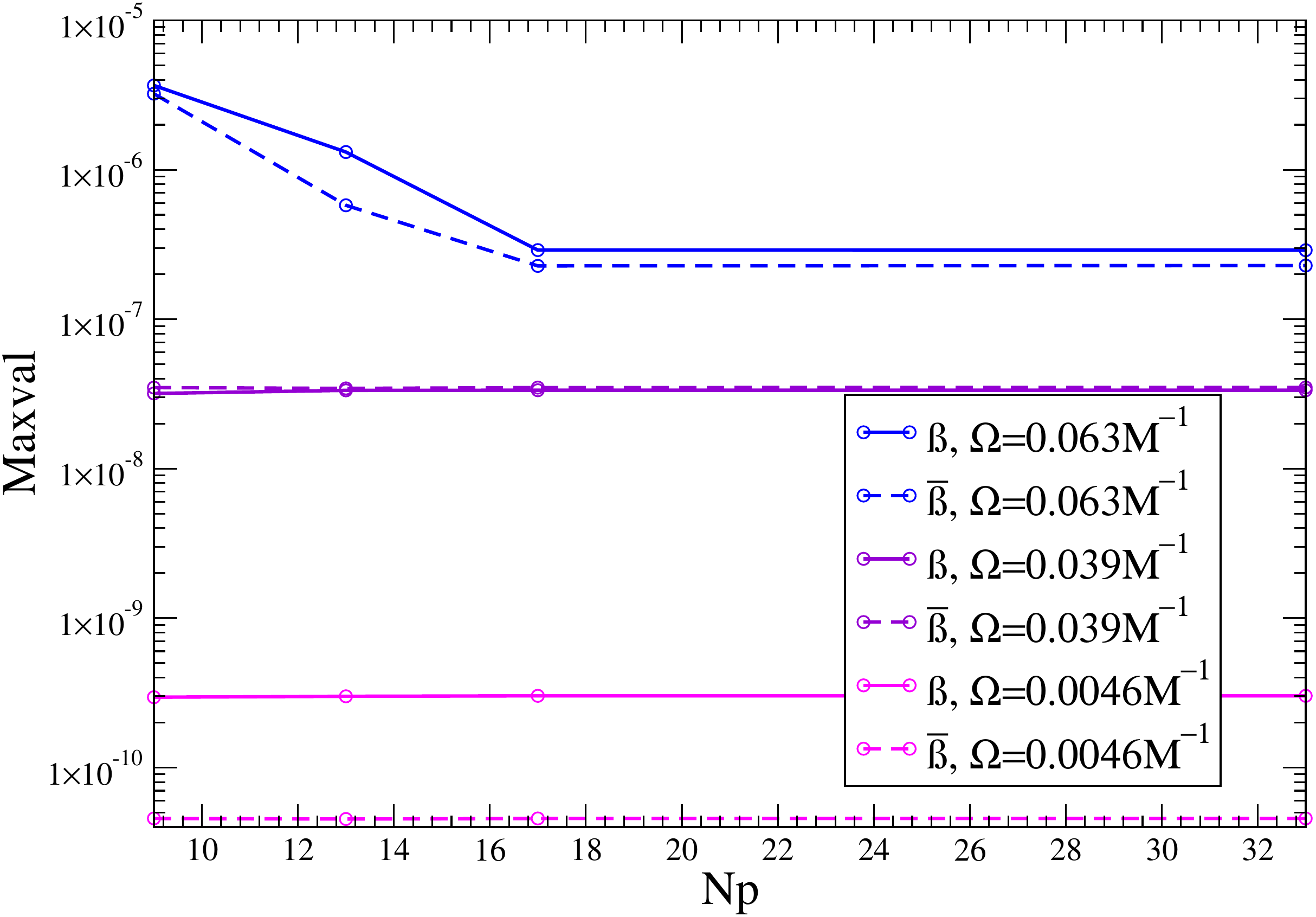}
\caption{Maximal values of $\ss$ and $\bar{\ss}$ as functions of the number
of spectral coefficients in both the radial and angular dimension
for neutron stars with for $\Omega=0.0046,0.039,0.063\: M^{-1}$ .
\label{fig:SpasiNSConvk}}
\end{figure}

We plot in Fig.~\ref{fig:SpasiNSConvk} the maximal values of $\ss$
and $\bar{\ss}$ for a static and several rotating neutron stars as
functions of the resolution.\\

As in the boson star case, the maximal values of $\ss$ and $\bar{\ss}$
do not depend on the resolution up to a certain precision. We choose
the resolution of $17$ points in $r$ and $\theta$ for the following
plots. \\

\begin{figure}[!hbtp]
\includegraphics[width=8cm]{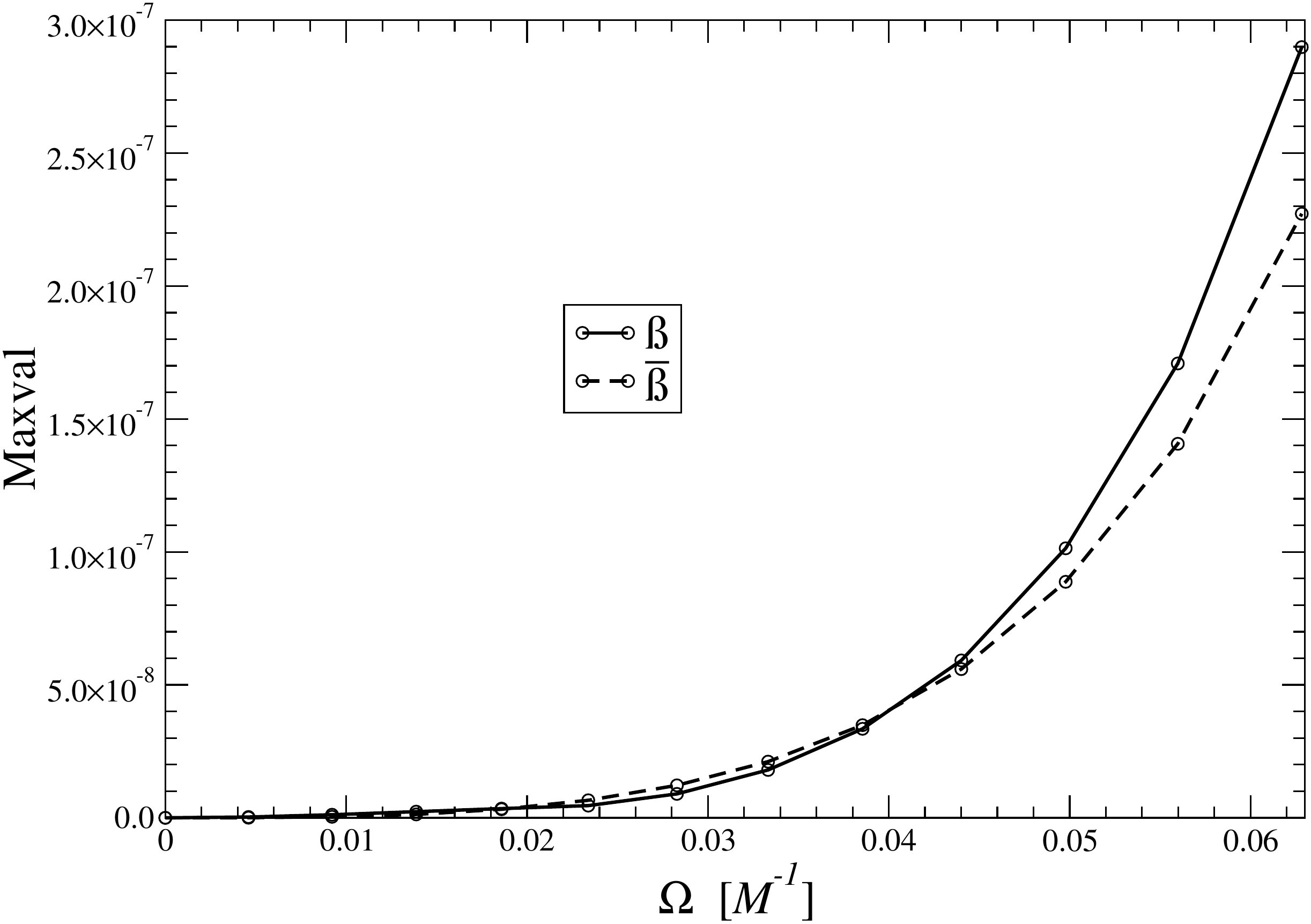}
\caption{Maximal values of $\ss$ and $\bar{\ss}$ as functions of $\Omega$
for rotating neutron stars. When $\Omega\rightarrow0$, the neutron
star is less rapidly rotating (static for $\Omega=0$) and the spacetime
becomes closer to the Schwarzschild solution.\label{fig:SpasiNSomega}}
\end{figure}

We can explore the global behavior of the maximal values of $\ss$
and $\bar{\ss}$ for different neutron stars for a fixed resolution.
For instance, we plot in Fig.~\ref{fig:SpasiNSomega} those maximal
values as functions of $\Omega$ for different rotating neutron stars.\\

\begin{figure}[!hbtp]
\includegraphics[width=4cm]{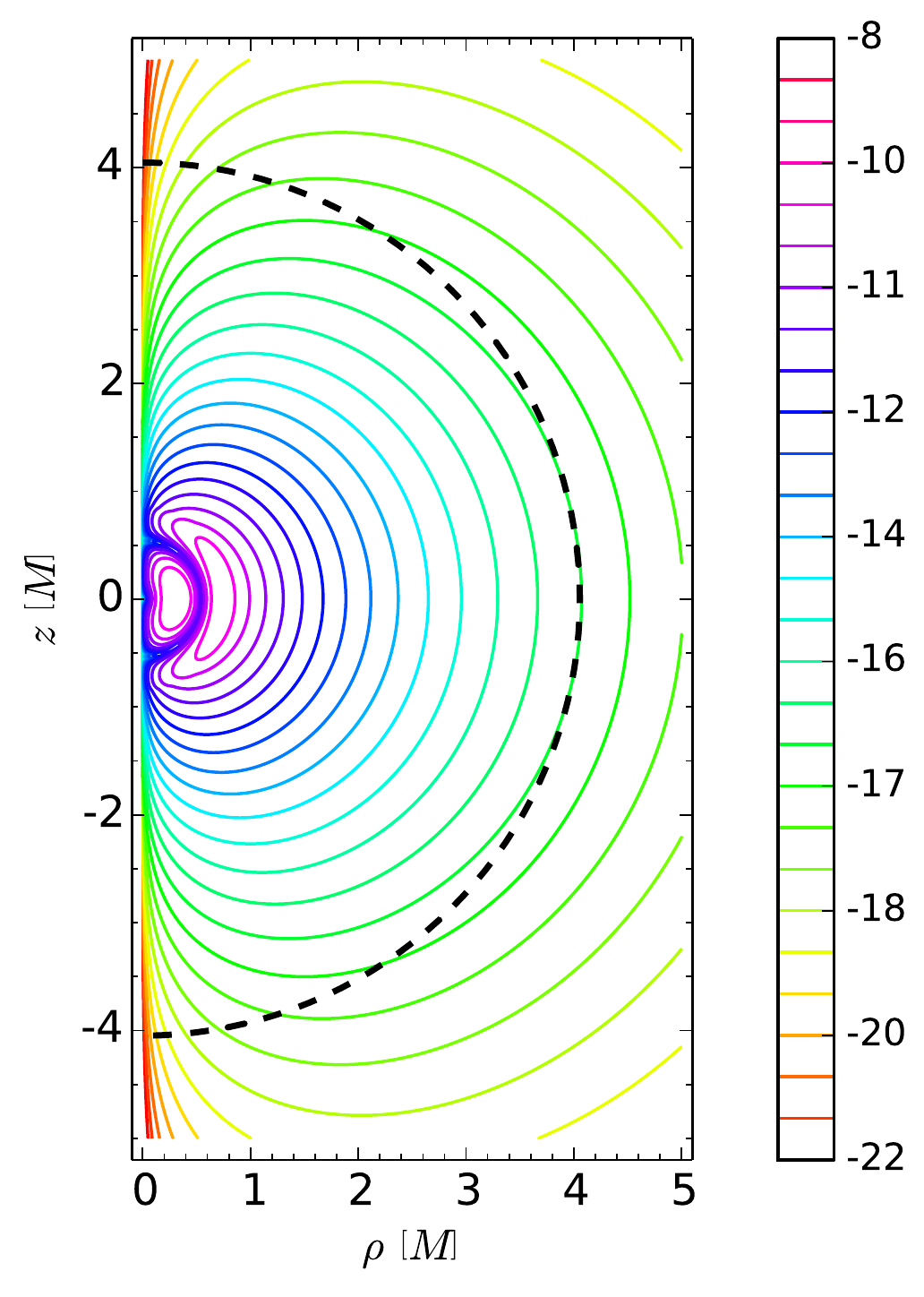}
\includegraphics[width=4cm]{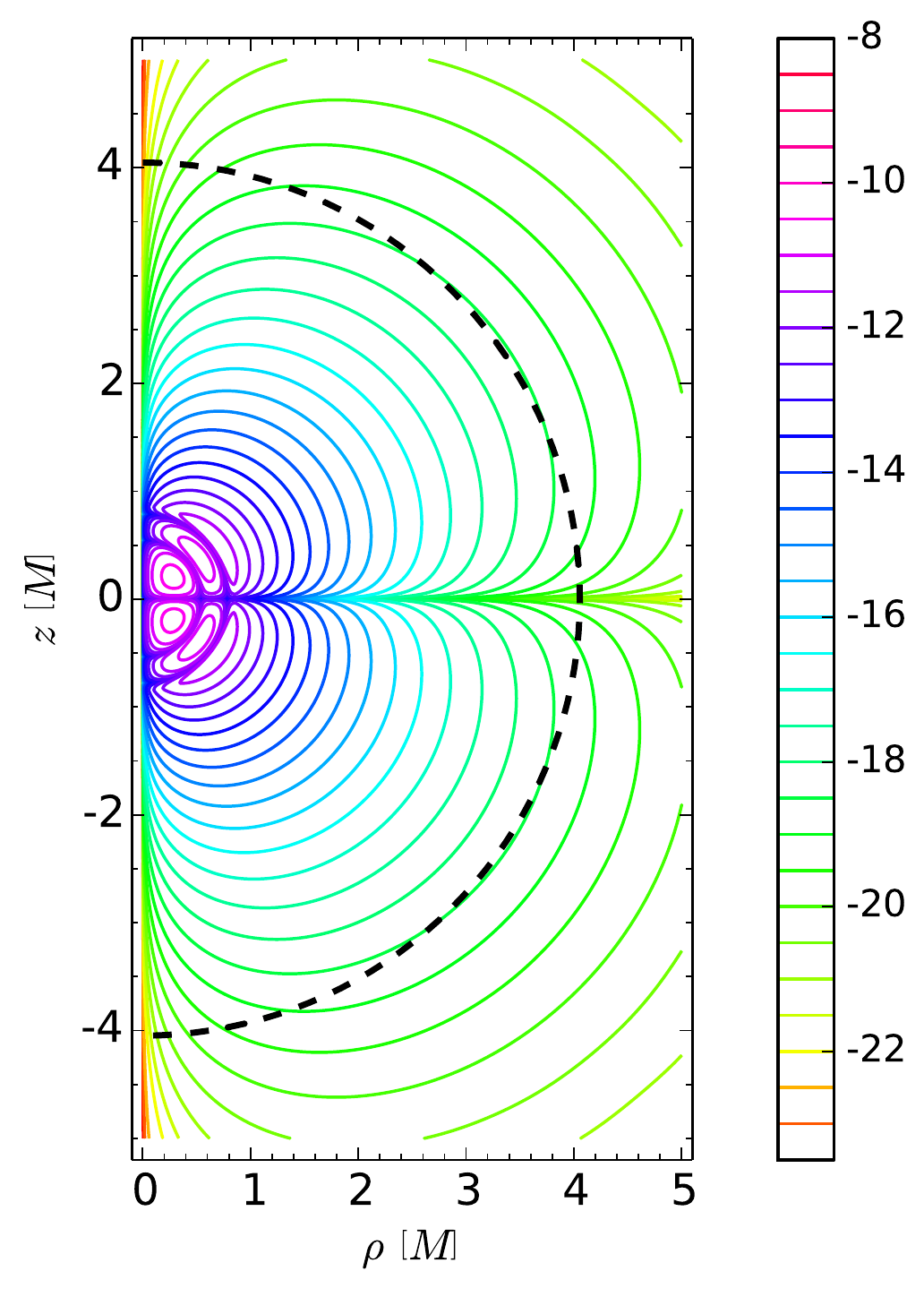}
\caption{Contour plot of $\log\left(\ss\right)$ as a function of $r$ and
$z$ for a rotating neutron star with $\Omega=0.0046\: M^{-1}$. The
black dashed line indicates the position of the surface of the neutron
star.\label{fig:SpasiNSR12}}
\end{figure}

\begin{figure}[!hbtp]
\includegraphics[width=8cm]{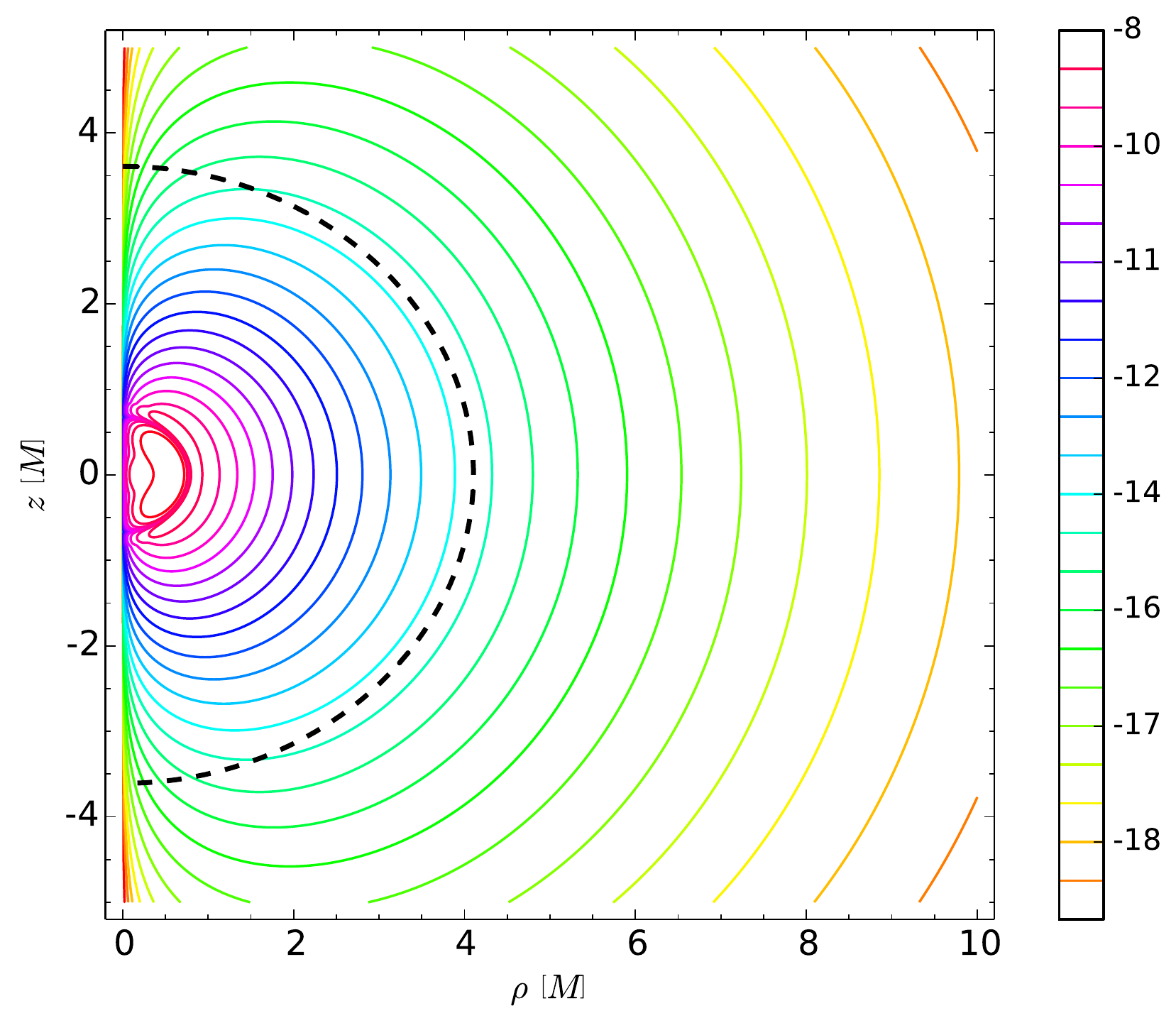}
\includegraphics[width=8cm]{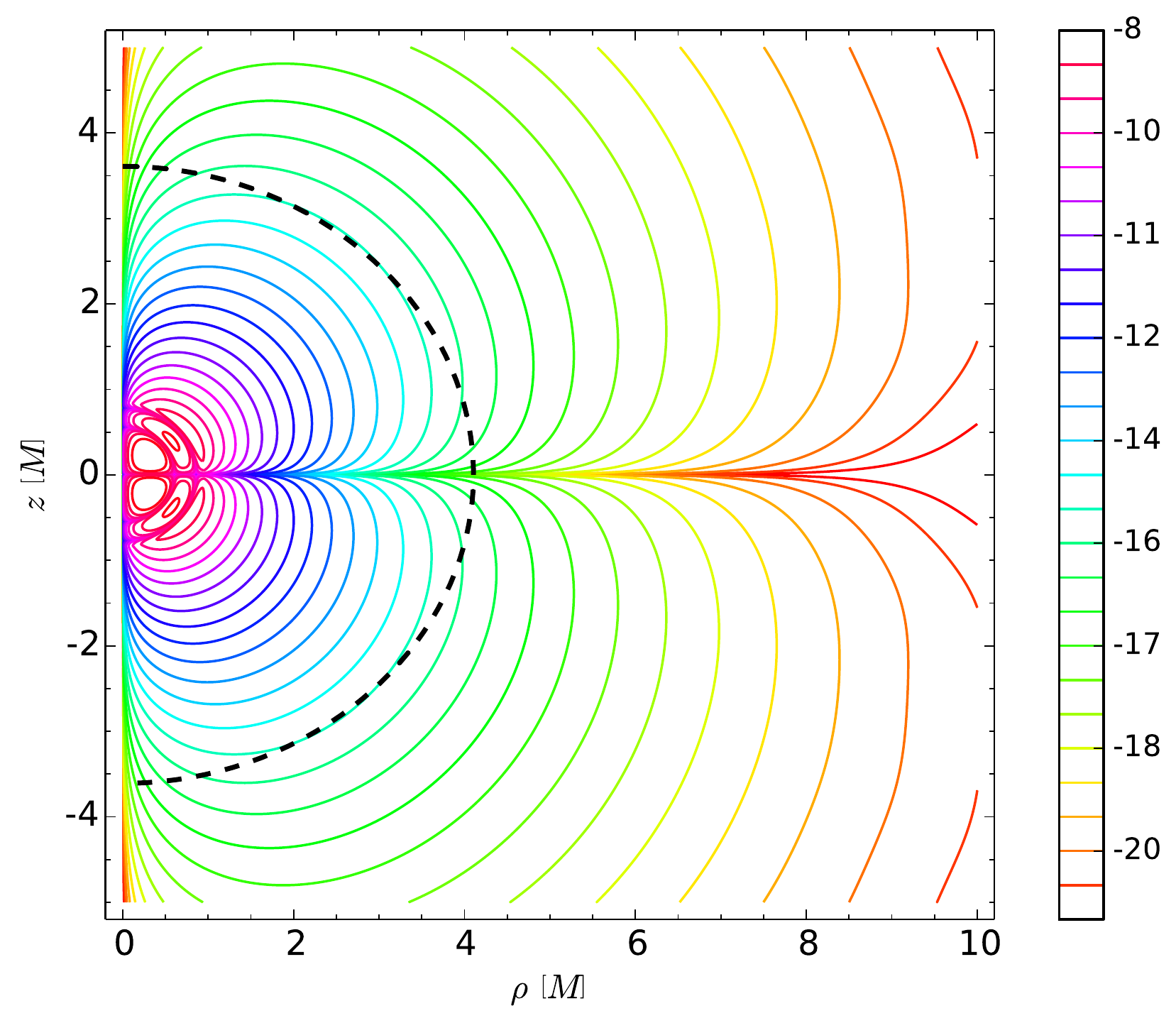}
\caption{Contour plots of $\log\left(\ss\right)$ and $\log\left(\bar{\ss}\right)$
as functions of $r$ and $z$ for a rotating neutron star with $\Omega=0.039\: M^{-1}$.\label{fig:SpasiNSR12-2}}
\end{figure}

Figures \ref{fig:SpasiNSR12} and \ref{fig:SpasiNSR12-2} 
show contours of $\log\left(\ss\right)$ and $\log\left(\bar{\ss}\right)$ 
for two different neutron stars.

\subsection{\label{sec:Analytical-applications}Exact analytic solutions of the
Einstein equations}

It is interesting also to compare the values of the Simon-Mars scalars
we obtain in the two preceding examples to other spacetimes. We chose
to do it with two specific spacetimes for which the difference to
Kerr has been already quantified in \cite{GLS13}, one is the Curzon-Chazy
spacetime first studied by Curzon in \cite{Curzon} and Chazy in \cite{Chazy}, the other
is the $\delta=2$ Tomimatsu-Sato spacetime defined in \cite{Tom72}.\\

Calculating the Simon-Mars scalars for these spacetimes permits us
to put a link between our characterization of the ``non-Kerness''
and the characterization with invariants given in \cite{GLS13}. This
classification is more fine than ours because is uses more than the
Simon-Mars tensor. But it is based on vacuum spacetime, or very small
deviation from vacuum spacetimes while ours can be applied to all
spacetimes. Thus, our scalar calculation can advantageously complete
their classification providing invariants that can be calculated even
for non vacuum spacetimes such as the two examples treated in the
preceding section. \\

Even if there are more examples in \cite{GLS13}, in most of them
(which are not vacuum solutions of the Einstein equations, so Mars
theorem does not hold), the Simon-Mars is zero, so we focus in the
solutions that are in vacuum and not locally isomorphic to Kerr (their
Simon-Mars tensor cannot be zero). For this part, we used the
SageManifolds free package \cite{GourgBM14,SageManifolds}, which is an
extension towards differential geometry and tensor calculus of the
open-source mathematics software Sage \cite{Sage}. 

\subsubsection{Exact vacuum axisymmetric solutions}

As we saw in section \ref{sub:Quasi-isotropic-coordinates}, the metric
of a generic stationary, axisymmetric and circular spacetime can be
written in quasi-isotropic coordinates (\ref{eq:metQI}) : $\left(t,r,\theta,\varphi\right)$.
The related cylindrical coordinates $\left(t,\rho,z,\varphi\right)$
with $\rho=r\sin\theta$ and $z=r\cos\theta$ are the Weyl-Lewis-Papapetrou
coordinates :
\begin{eqnarray}
\textrm{d}s^{2} & = & -e^{2U}\left(\textrm{d}t+\beta^{\varphi}\rho\textrm{d}\varphi\right)^{2}\nonumber \\
 &  & +e^{-2U}\left[e^{2\gamma}\left(\textrm{d}\rho^{2}+\textrm{d}z^{2}\right)+\rho^{2}\textrm{d}\varphi^{2}\right],\label{eq:WLPmet}
\end{eqnarray}
where $U$ and $\gamma$ are functions of $\rho$ and $z$. In these
coordinates, $U$ and $\gamma$ play the same role of $A,\, B$ and
$N$ in the quasi-isotropic coordinates : indeed, in vacuum, $B=1/N$
(see eq. (3.16) of \cite{Rotstar}). Writing the vacuum Einstein field
equations satisfied by this metric, we obtain a single differential
equation, called the Ernst equation, for a complex potential, called
the Ernst potential \cite{Ernst}. This equation is integrable so
it can be solved by various solutions generating techniques : in the
static case ($\beta^{\varphi}=0$), a family of asymptotically flat
solutions can be found expressing $U$ as a sum of Legendre polynomials.
This family is called the Weyl class \cite{GP} and Curzon-Chazy spacetime
is the simplest member of this class. \\

For stationary (but not static) axisymmetric solutions, it is convenient
to write the Ernst equation in prolate spherical coordinates $\left(x,y\right)$
related to the Weyl-Lewis-Papapetrou coordinates $\left(\rho,z\right)$
by 
\begin{eqnarray}
x & = & \frac{1}{2}\left(\sqrt{\rho^{2}+\left(z+1\right)^{2}}+\sqrt{\rho^{2}+\left(z-1\right)^{2}}\right)\label{eq:xPSC}\\
y & = & \frac{1}{2}\left(\sqrt{\rho^{2}+\left(z+1\right)^{2}}-\sqrt{\rho^{2}+\left(z-1\right)^{2}}\right),\label{eq:yPSC}
\end{eqnarray}
where $x\geq1$ and $y\in\left[-1,1\right]$. Tomimatsu and Sato \cite{Tom72, Sato73}
have labeled the solutions by a deformation parameter called $\delta$.
The solution corresponding to $\delta=1$, the simplest one, is the
Kerr one. The solution which we consider in this paper is the $\delta=2$
Tomimatsu-Sato solution.

\subsubsection{Kerr spacetime}

As a test of the SageManifolds code, we evaluated the Simon-Mars tensor
(\ref{eq:S}) for the Kerr metric in Boyer-Lindquist coordinates and
we evaluated also each of the eight 3+1 components of the Simon-Mars
tensor : (\ref{eq:R1})-(\ref{eq:R4}) and (\ref{eq:Rb1})-(\ref{eq:Rb4}).
We found identically zero. This computations confirms that the same
results are found with the original definition and with the 3+1 decomposition.
Furthermore, it validates the use of SageManifolds worksheets.
The latter are freely downloadable from \cite{SM_examples}. 

\subsubsection{Curzon-Chazy spacetime}

\begin{figure}[!hbtp]
\includegraphics[width=8cm]{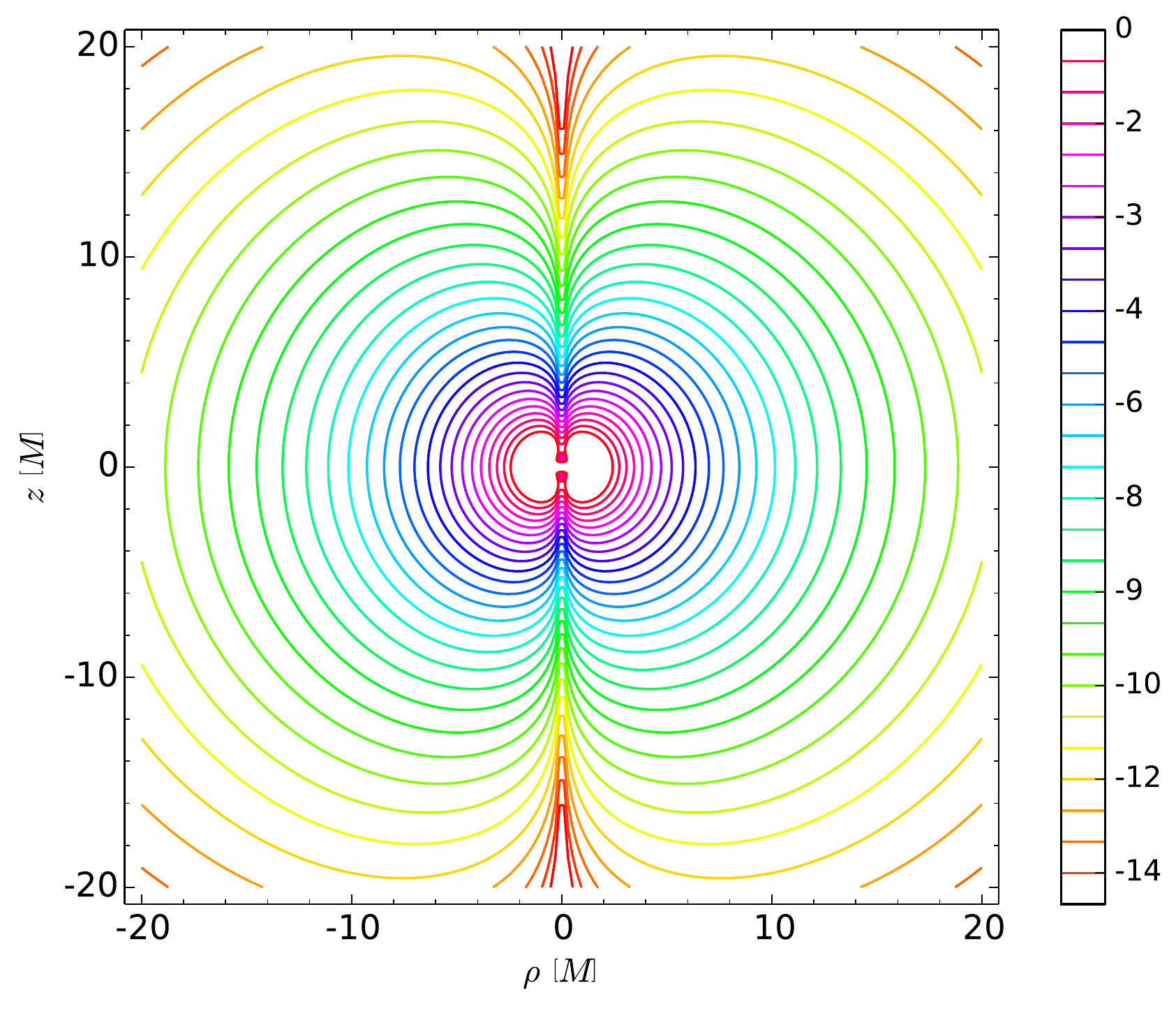}
\includegraphics[width=8cm]{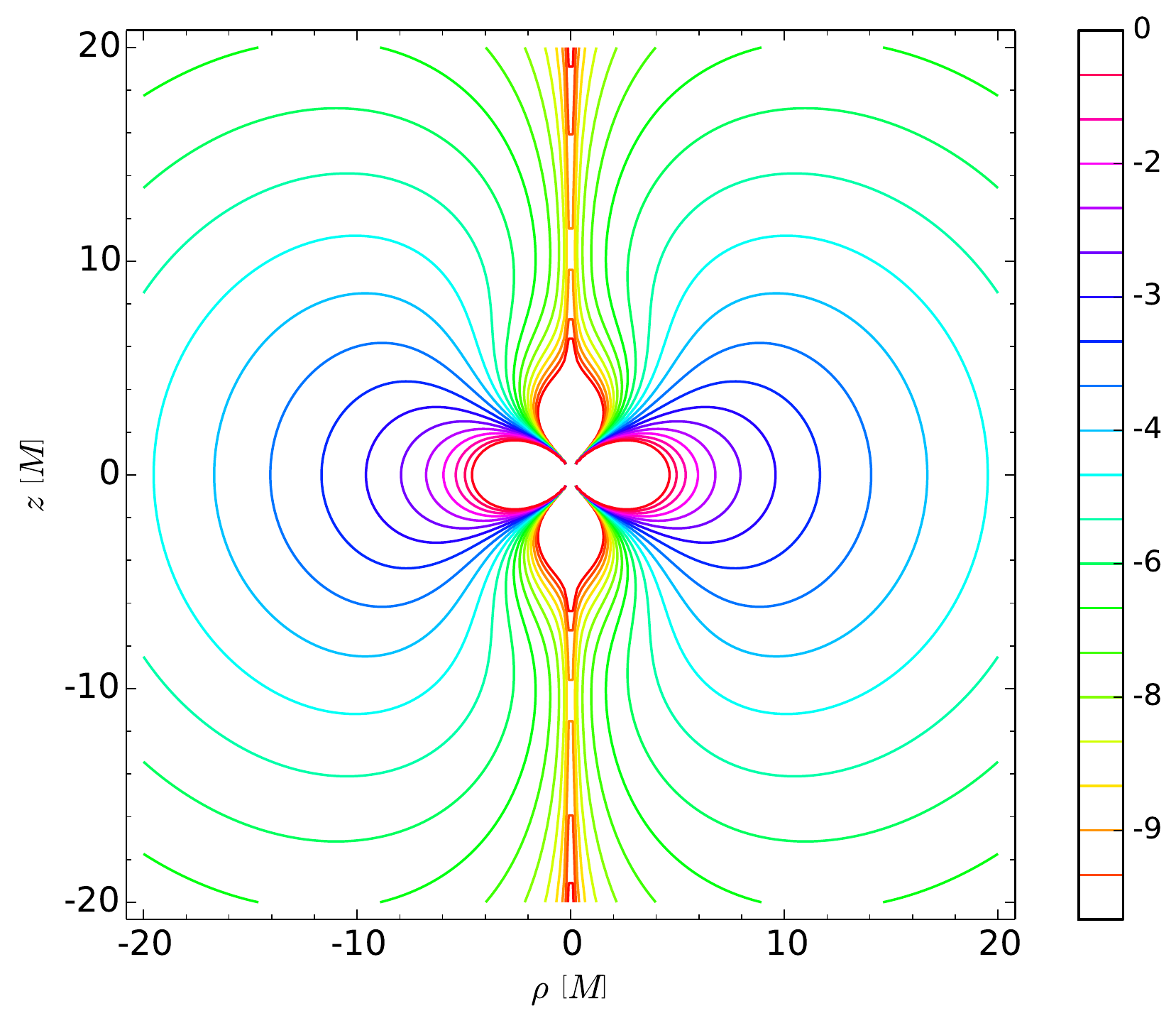}
\caption{Contour plots of $\log\left(\ss\right)$ for Curzon-Chazy spacetime
as a function of $\rho$ and $z$ for $M=1$ and for $M=12$.\label{fig:SpasiCCR12}}
\end{figure}

The Curzon-Chazy spacetime is a static, axisymmetric and asymptotically
flat solution of the Einstein equations \cite{Curzon,Chazy}. Its line element is given
by the Weyl-Lewis-Papapetrou one with $\beta^{\varphi}=0$ and 
\begin{equation}
U=-\frac{M}{r},\;\gamma=-\frac{M^{2}\sin^{2}\theta}{2r^{2}}.\label{eq:CCS}
\end{equation}
It has a spherically symmetric Newtonian potential, corresponding
to the potential of a point particle of mass $M$ located at $r=0$,
where lies a singularity of complex nature, but this spacetime is
not spherically symmetric. Because the shift is zero for this spacetime,
$\bar{\ss}$ is identically null, so we can only show the contour
plot of $\log\left(\ss\right)$ in Fig.~\ref{fig:SpasiCCR12}.

The computation of the Simon-Mars scalars has been performed with
SageManifolds (the corresponding worksheet is available at
\cite{SM_examples}). As this solution is analytic, we identified those scalars
both with their 3+1 definitions (\ref{eq:1/4ReS}) and (\ref{eq:1/4ImS}),
but also as the real and imaginary part of the ``square'' of the
4-dimensional Simon-Mars scalar (\ref{eq:Square}). \\

First we see that the Simon-Mars scalars diverge at the singularity.
For vacuum spacetimes, the Weyl tensor is equal to the Riemann tensor,
so the Simon-Mars are close to the Kretchmann with diverge at each
real singularity (not coordinate singularities) of the metric, so
we can expect this behavior. Furthermore, comparing with Fig.~1
of \cite{GLS13}, we can confirm that the adapted quantity is $\log\left(\ss\right)$
and not $\ss$ by comparing the shape of the contours, we can also
see that a spherically symmetric spacetime differs little to the Kerr
spacetime when $\log\left(\ss\right)<-8$. In this sense, at the singularity
the Curzon-Chazy spacetime is infinitely far from the Kerr spacetime,
which is not false.

\subsubsection{$\delta=2$ Tomimatsu-Sato spacetime}

This spacetime, which is defined in \cite{Tom72,Sato73}, is a
stationary, axisymmetric vacuum solution of the Einstein equations
which is also asymptotically flat (see \cite{Glass73}). This metric
is one of the rare stationary and axisymmetric exact solutions of
Einstein equations. The non zero elements of the metric are 
\begin{eqnarray}
g_{tt} & = & -\frac{A\left(x,y\right)}{B\left(x,y\right)},\label{eq:gtt}\\
g_{t\varphi} & = & -\frac{2Mq\left(x,y\right)C\left(x,y\right)\left(1-y^{2}\right)}{B\left(x,y\right)},\label{eq:gtp}\\
g_{xx} & = & \frac{M^{2}B\left(x,y\right)}{p^{2}\delta^{2}\left(x^{2}-1\right)\left(x^{2}-y^{2}\right)^{3}},\label{eq:gxx}\\
g_{yy} & = & \frac{M^{2}B\left(x,y\right)}{p^{2}\delta^{2}\left(y^{2}-1\right)\left(y^{2}-x^{2}\right)^{3}},\label{eq:gyy}\\
g_{\varphi\varphi} & = & \frac{M^{2}\left(y^{2}-1\right)p^{2}B^{2}\left(x,y\right)\left(x^{2}-1\right)}{A\left(x,y\right)B\left(x,y\right)\delta^{2}}\nonumber \\
 &  & +4\frac{q^{2}\delta^{2}C^{2}\left(x,y\right)\left(y^{2}-1\right)}{A\left(x,y\right)B\left(x,y\right)\delta^{2}},\label{eq:gpp}
\end{eqnarray}
with 
\begin{eqnarray}
A\left(x,y\right) & = & \left[p^{2}\left(x^{2}-1\right)^{2}+q^{2}\left(1-y^{2}\right)^{2}\right]^{2}\nonumber \\
 &  & -4p^{2}q^{2}\left(x^{2}-1\right)\left(1-y^{2}\right)\left(x^{2}-y^{2}\right)^{2},\label{eq:ATS}\\
B\left(x,y\right) & = & \left(p^{2}x^{4}+q^{2}y^{4}-1+2px^{3}-2px\right)^{2}\nonumber \\
 &  & +4q^{2}y^{2}\left(px^{3}-pxy^{2}+1-y^{2}\right)^{2},\label{eq:BTS}\\
C\left(x,y\right) & = & p^{2}\left(x^{2}-1\right)\left[\left(x^{2}-1\right)\left(1-y^{2}\right)-4x^{2}\left(x^{2}-y^{2}\right)\right]\nonumber \\
 &  & -p^{3}x\left(x^{2}-1\right)\left[2\left(x^{4}-1\right)+\left(x^{2}+3\right)\left(1-y^{2}\right)\right]\nonumber \\
 &  & +q^{2}\left(1+px\right)\left(1-y^{2}\right)^{3},\label{eq:CTSb}
\end{eqnarray}
where $p$ and $q$ are real constant satisfying the constraint $p^{2}+q^{2}=1$,
and $M$ is the ADM mass, here we choose $p=1/5$%
\footnote{same as in \cite{GLS13}%
}, and $M=1$. \\

This spacetime contains two degenerated Killing horizons at $x=1$,
$y=\pm1$, and a naked ring singularity (more on the $\delta=2$ Tomimatsu-Sato
spacetime in \cite{Manko}). The computation of the two Simon-Mars scalars
starting from the metric (\ref{eq:gtt})-(\ref{eq:CTSb}) 
is rather formidable. We performed it by means of the SageManifolds code
mentioned above. The corresponding worksheet is available at
\cite{SM_examples}.

\begin{figure}[!hbtp]
\includegraphics[width=4cm]{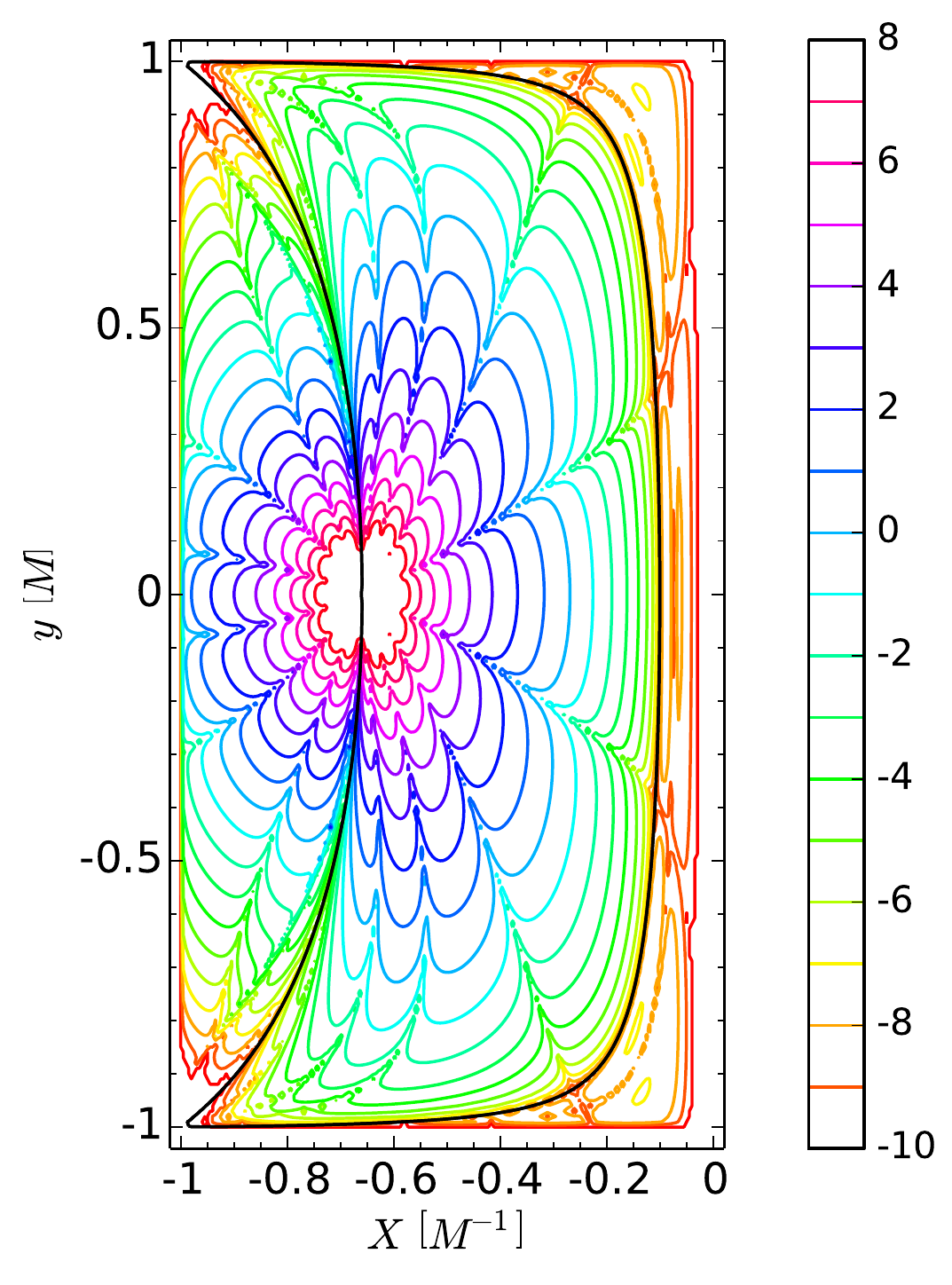}
\includegraphics[width=4cm]{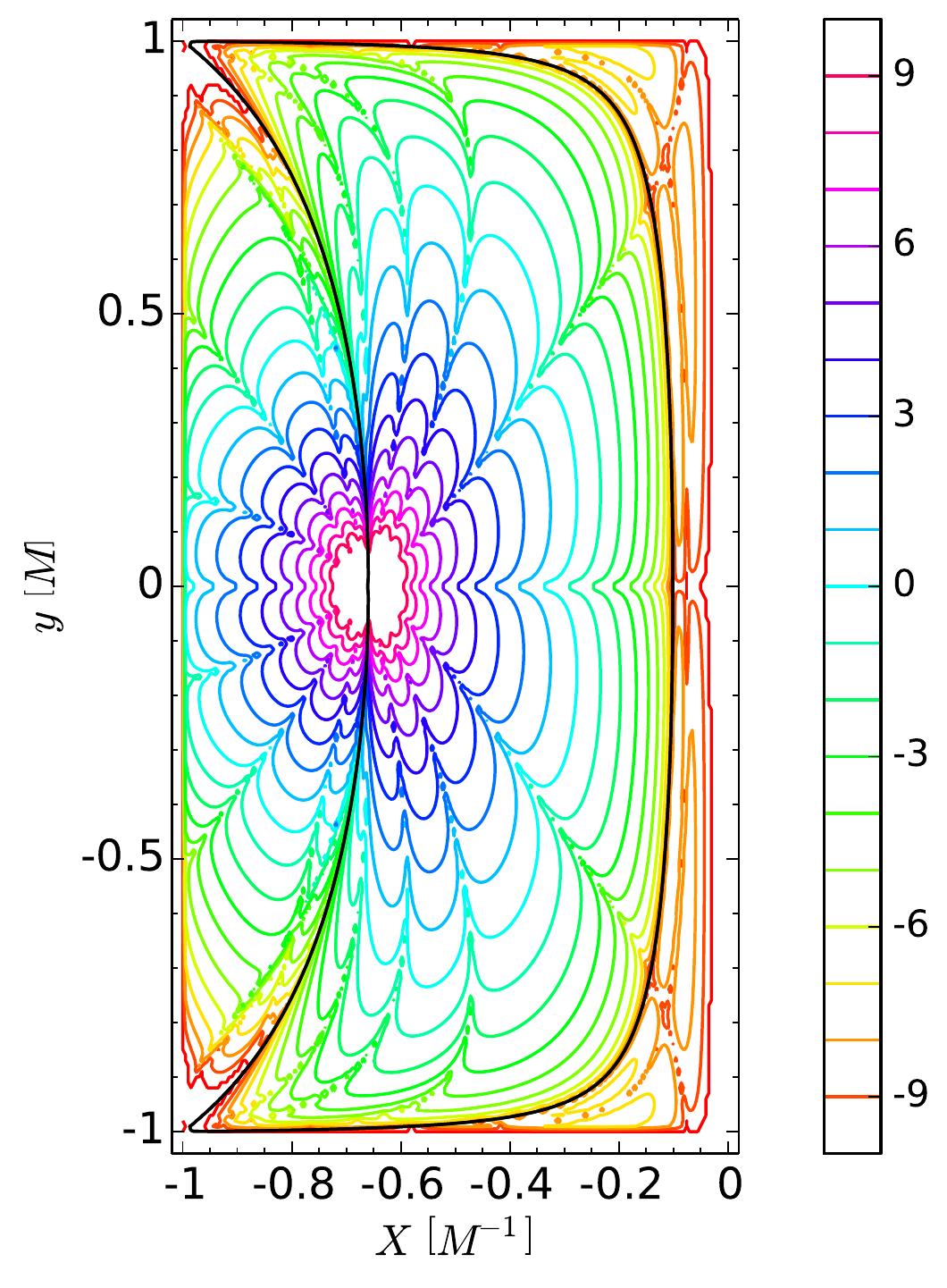}
\caption{Contour plots of $\log\left(\ss\right)$ and $\log\left(\bar{\ss}\right)$
for $\delta=2$ Tomimatsu-Sato spacetime for $p=1/5$ and $M=1$ and
as functions of $X=-1/x$ and $y$. The black line marks the ergosurface.
The scalars diverge at the naked ring singularity. \label{fig:SpasiTSR12}}
\end{figure}

In Fig.~\ref{fig:SpasiTSR12}, we show
the contour plots of $\log\left(\ss\right)$ and $\log\left(\bar{\ss}\right)$
as functions of $X=-1/x$ and $y$ (given by (\ref{eq:xPSC}) and
(\ref{eq:yPSC})) to compare our result to the Fig.~2 of \cite{GLS13}.
We see that the order of magnitude of the two scalars is the same.
But is is harder to make a comparison with Fig.~2 of \cite{GLS13}
in this case because the scalars grow exponentially around the singularity,
which is not the case in \cite{GLS13}. Nevertheless we can compare
the value of the scalars along the horizontal axis ($y=0$), and we
see that in this case an axisymmetric spacetime seems to differ little
to the Kerr spacetime when $\log\left(\ss\right)<-6$. Outside the
ergosphere the values of $\log\left(\ss\right)$ and $\log\left(\bar{\ss}\right)$
diminish frankly. \\

\begin{figure}[!hbtp]
\includegraphics[width=8cm]{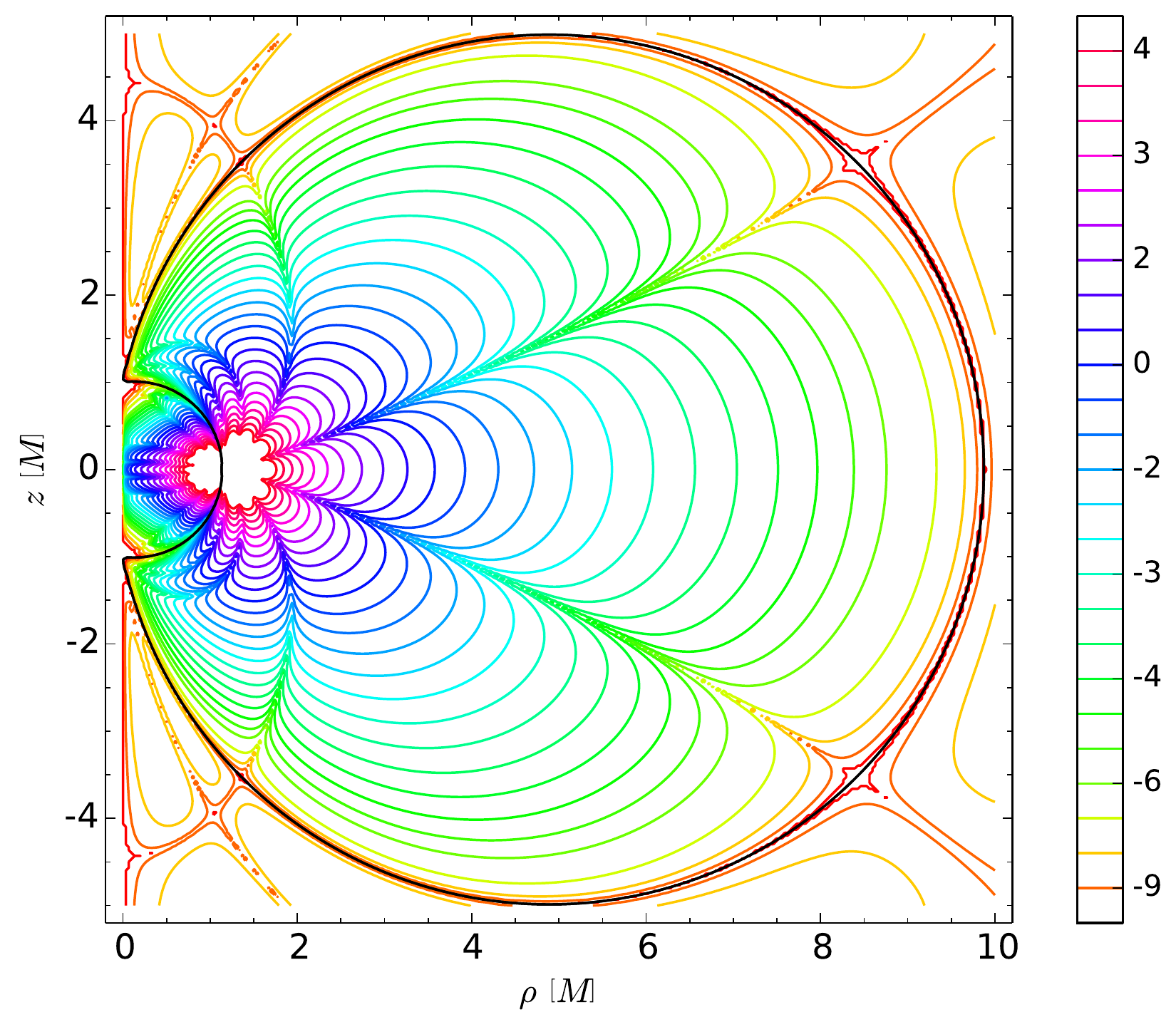}
\includegraphics[width=8cm]{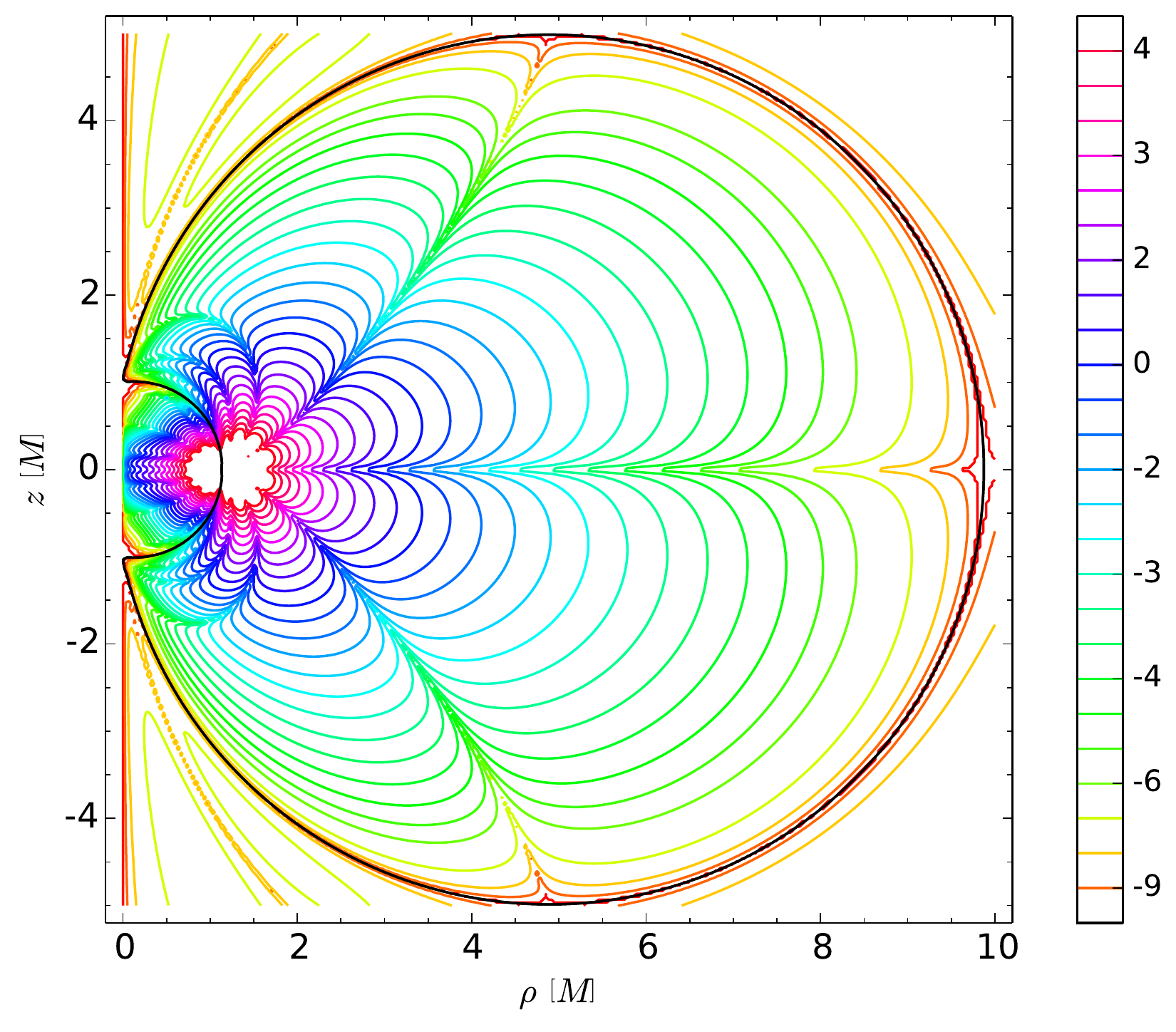}
\caption{Same plot as in Fig.~\ref{fig:SpasiTSR12} in Weyl-Lewis-Papapetrou
coordinates $\rho$ and $z$. \label{fig:SpasiTSR12-1}}
\end{figure}

We do the same plot in the Weyl-Lewis-Papapetrou coordinates which
are more usual coordinates (see \cite{Manko}), in Fig.~\ref{fig:SpasiTSR12-1}. 

\subsection{Approximate solution : the Modified Kerr metric}

The modified Kerr metric proposed by Johannsen and Psaltis \cite{JoPs}
is a family of approximate solutions of the Einstein field equations,
which are non linear parametric deviations from the Kerr metric. The
line element is given in Boyer-Lindquist coordinates by 
\begin{eqnarray}
\textrm{d}s^{2} & = & -\left[1+h\left(r,\theta\right)\right]\left(1-\frac{2Mr}{\Sigma}\right)\textrm{d}t^{2}-\frac{4aMr\sin^{2}\theta}{\Sigma}\nonumber \\
 &  & \times\left[1+h\left(r,\theta\right)\right]\textrm{d}t\textrm{d}\varphi+\frac{\Sigma\left[1+h\left(r,\theta\right)\right]}{\Delta+a^{2}\sin^{2}\theta h\left(r,\theta\right)}\textrm{d}r^{2}\nonumber \\
 &  & +\Sigma\textrm{d}\theta^{2}+\left[\sin^{2}\theta\left(r^{2}+a^{2}+\frac{2a^{2}Mr\sin^{2}\theta}{\Sigma}\right)\right.\nonumber \\
 &  & +\left.h\left(r,\theta\right)\frac{a^{2}\left(\Sigma+2Mr\right)\sin^{4}\theta}{\Sigma}\right]\textrm{d}\varphi^{2},\label{eq:MKmet}
\end{eqnarray}
where $a$ is the specific angular momentum of the black hole, $M$
its ADM mass and 
\begin{eqnarray}
\Sigma & = & r^{2}+a^{2}\cos^{2}\theta\label{eq:sigm}\\
\Delta & = & r^{2}-2Mr+a^{2}.\label{eq:delt}
\end{eqnarray}
The simplest choice for $h\left(r,\theta\right)$ in accordance with
the observational constraints on weak-field deviations from general
relativity is given by
\begin{equation}
h\left(r,\theta\right)=\epsilon_{3}\frac{M^{3}r}{\Sigma^{2}}.\label{eq:h}
\end{equation}
This spacetime is a black hole only for certain values of $\epsilon_{3}$
and $a$ (see Fig.~3 of \cite{Jo13}) : if $0\leq a\leq0.8\: M$
and $-5\leq\epsilon_{3}\leq0$, it is always the case. In this paper,
we focus on this particular zone, so we use the positive parameter
$\epsilon=-\epsilon_{3}$. \\

\begin{figure}[!hbtp]
\includegraphics[width=8cm]{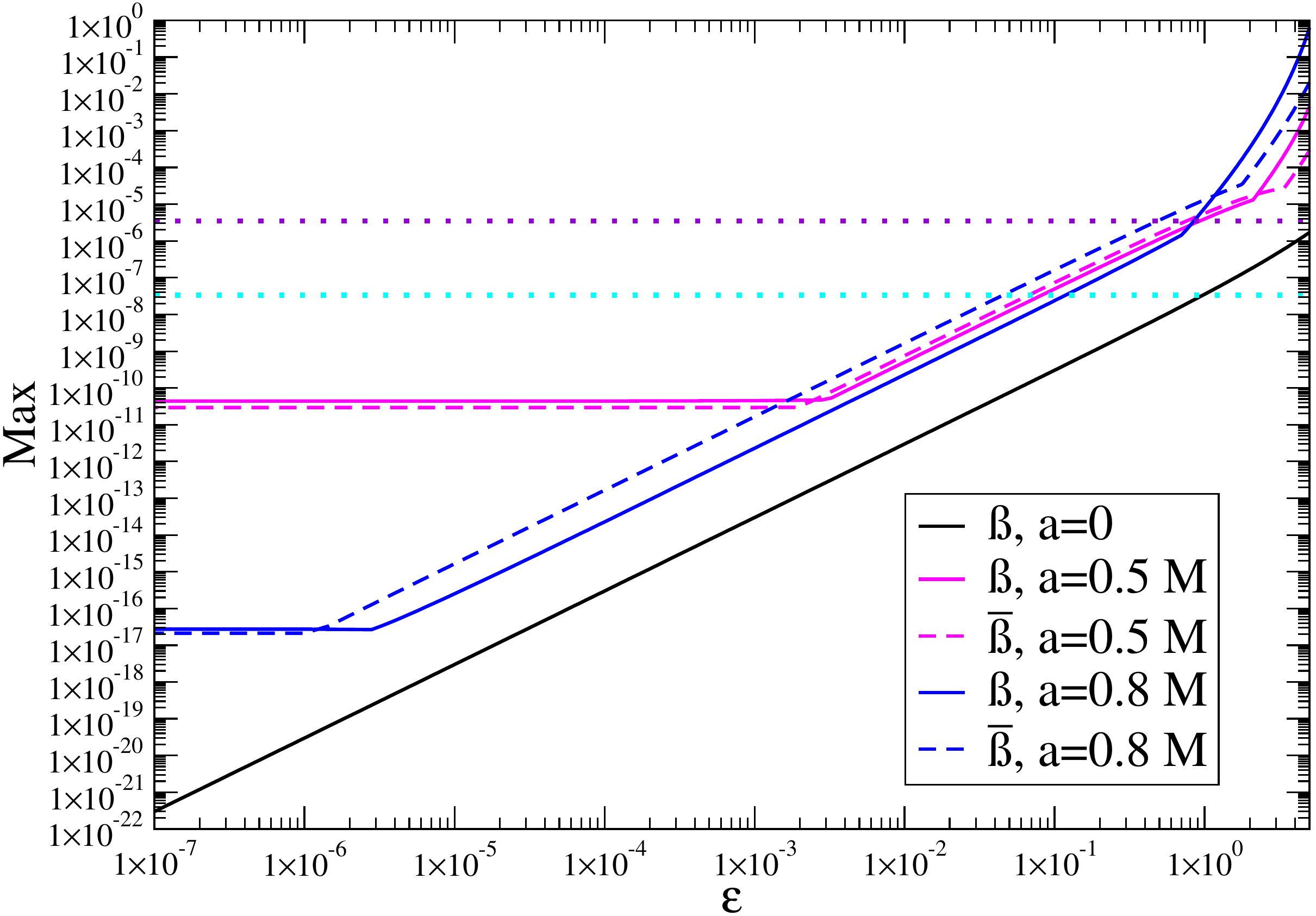}
\caption{Maximal values $\ss$ and $\bar{\ss}$ as functions of $\epsilon$
for modified Kerr metrics with $a=0,0.5,0.8\: M$. As the deviation
from the Kerr metric is non linear, we use a log-log scale. The violet
dotted line corresponds to the maximal value of $\ss$ ($\ss=3.48826.10^{-6}$)
for the boson star with $\omega=1.05\, M^{-1}$ and $k=1$, and the
cyan dotted line corresponds to the maximal value of $\ss$ ($\ss=3.34572.10^{-8}$)
for the neutron star with $\Omega=0.039\: M^{-1}$. \label{fig:SpasiMKeps}}
\end{figure}

In Fig.~\ref{fig:SpasiMKeps}, we plot the maximal values of $\ss$
and $\bar{\ss}$ as functions of $\varepsilon$ for different values
of the spin. We see that the bigger $a$ is, the bigger is the lowest
maximal value of the scalars, meaning that the more this modified
Kerr black hole is rapidly rotating, the more it differs for the Kerr
original solution. The idea was to quantify the deviation to the Kerr
spacetime of the preceding examples by seeing how fast the two Simon-Mars
scalars evaluated in the modified Kerr metric deviate from zero. \\

For spacetimes containing singularities as the Curzon-Chazy solution
and the $\delta=2$ Tomimatsu-Sato spacetime, the characterization
of their ``non-Kerness'' can be only local because the two scalars
diverge at the singularities. But for smooth spacetimes such as boson
stars or neutron stars, we can report the maximal values of the scalars
on the Fig.~\ref{fig:SpasiMKeps}, as we did for one specific example
of each. We can say that the spacetime generated by the neutron star
chosen is more close to the Kerr spacetime than the spacetime generated
by the boson star. This corresponds to our intuition because the boson
star shape is a torus, and it seems a more exotic object than the
neutron star.\\

\begin{figure}[!hbtp]
\includegraphics[width=8cm]{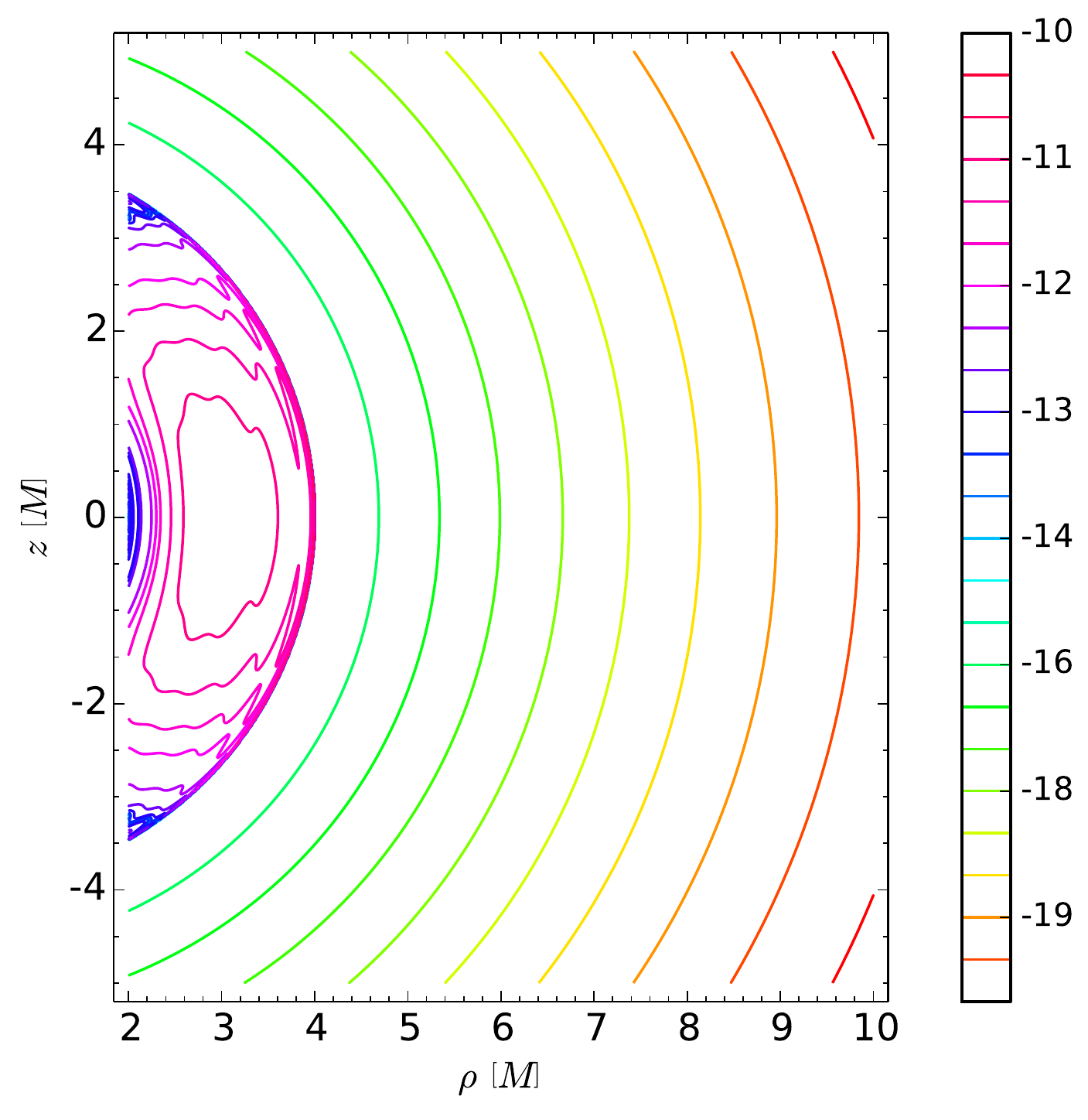}
\includegraphics[width=8cm]{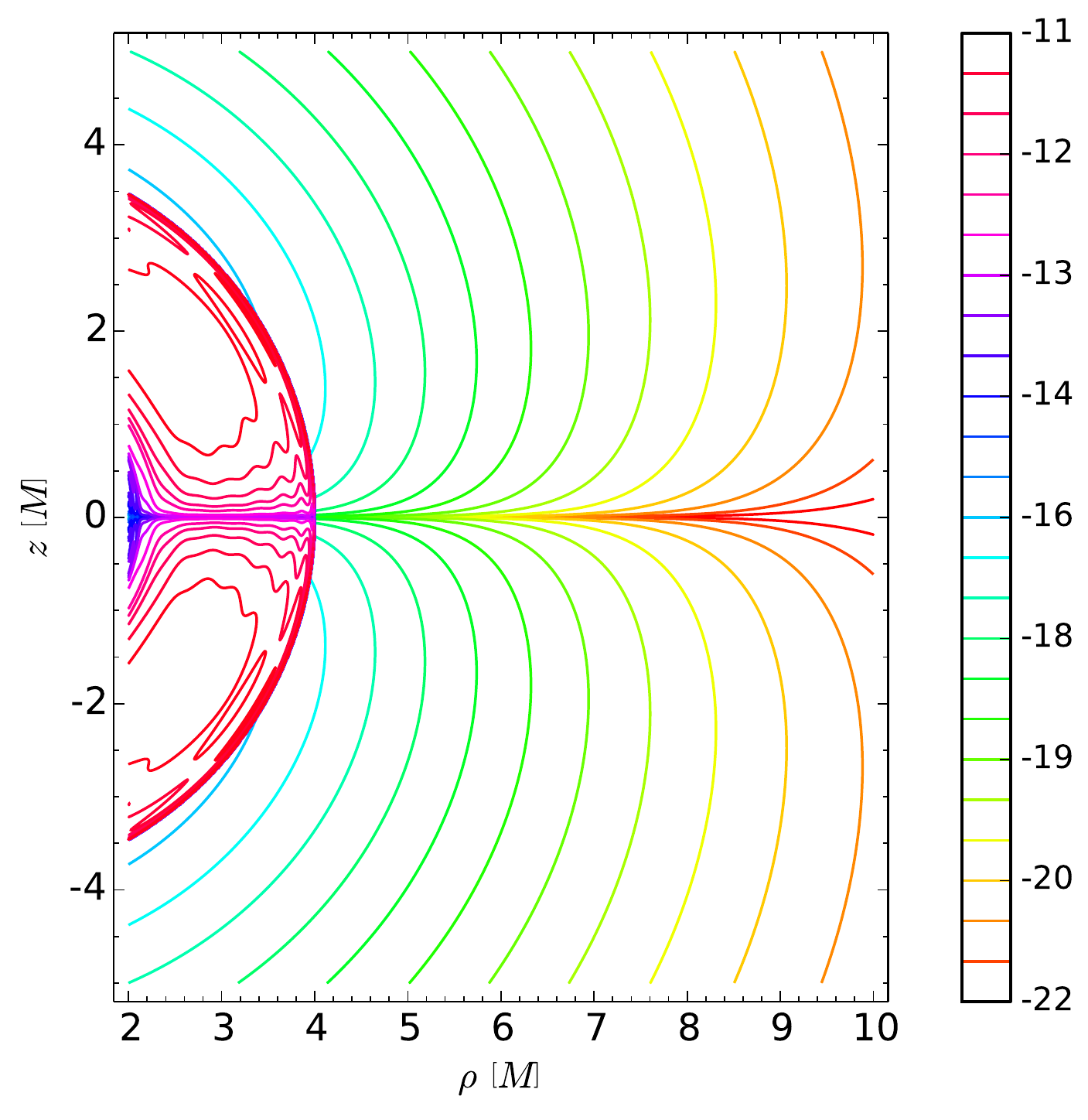}
\caption{Contour plot of $\log\left(\ss\right)$ and $\log\left(\bar{\ss}\right)$
as functions of $r\geq2\, M$ and $z$ for a modified Kerr metric
with $a=0.5\: M$ and $\epsilon=10^{-4}$.\label{fig:SpasiNSR12-1}}
\end{figure}

\begin{figure}[!hbtp]
\includegraphics[width=8cm]{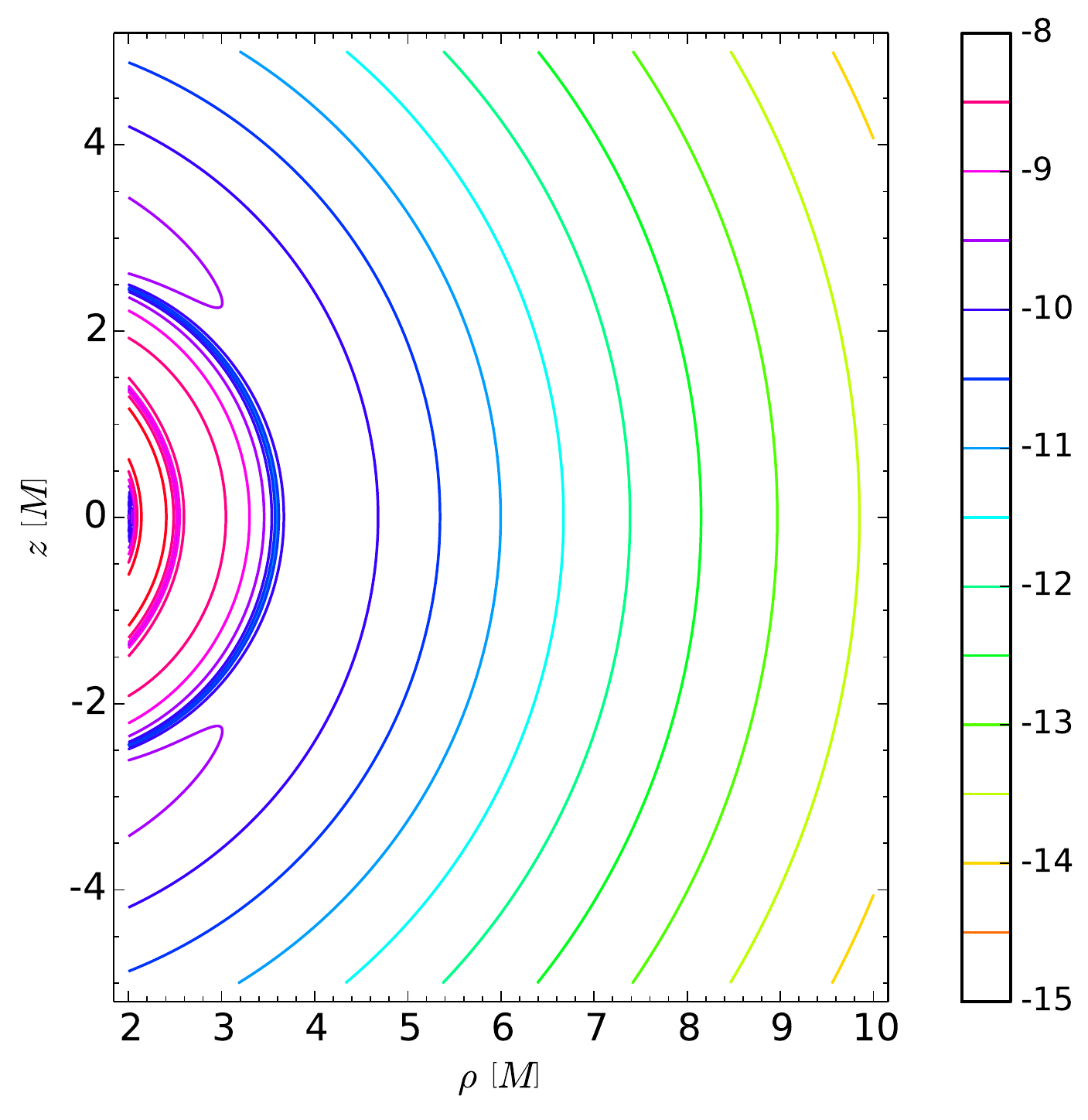}
\includegraphics[width=8cm]{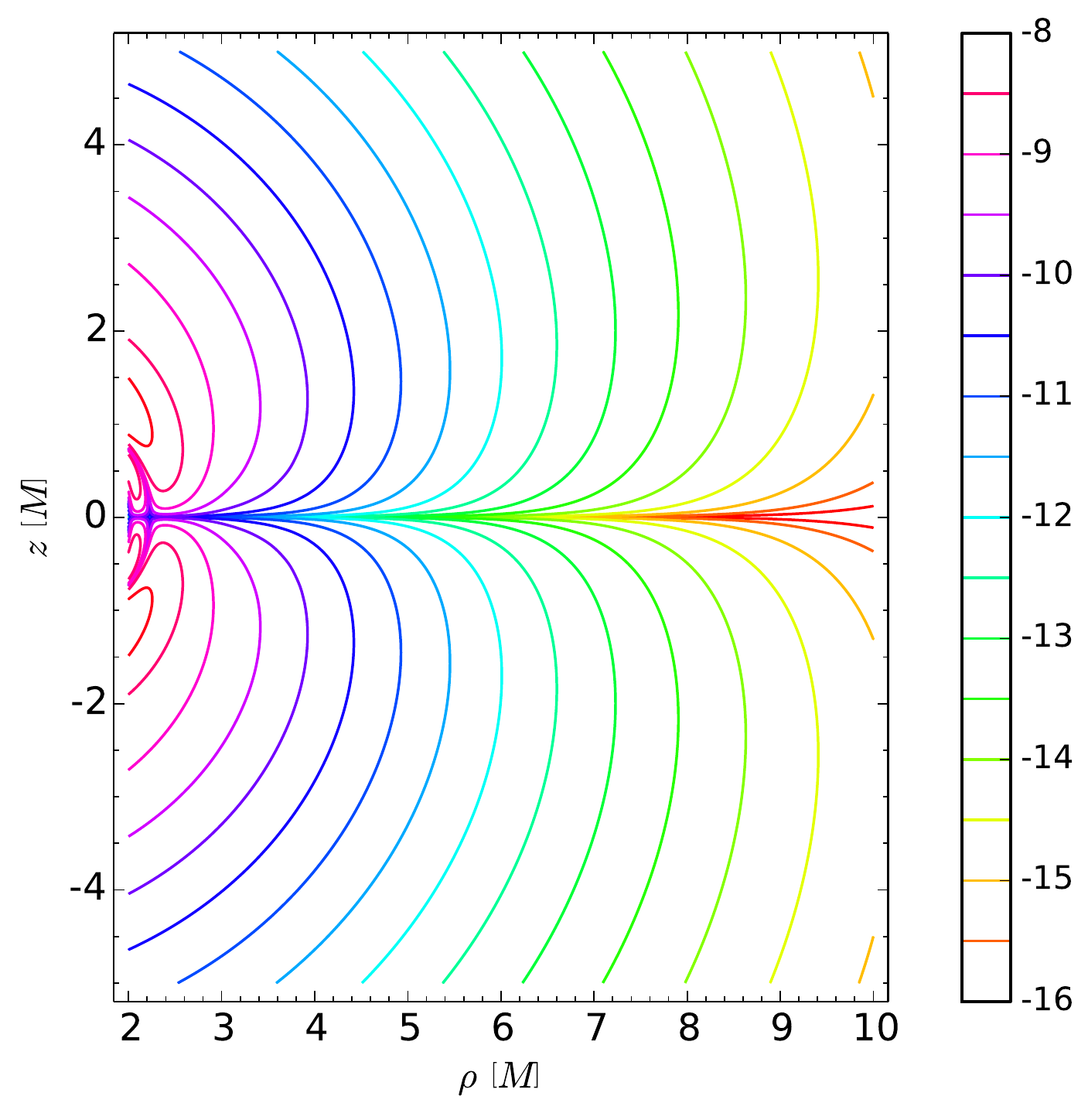}
\caption{Same plots as Fig.~\ref{fig:SpasiNSR12-1} a modified Kerr metric
with $a=0.8\: M$ and $\varepsilon=0.1$.\label{fig:SpasiNSR12-1-1}}
\end{figure}

To see local configurations of the scalar fields for examples of modified
Kerr spacetimes, we present contour plots of $\log\left(\ss\right)$
and $\log\left(\bar{\ss}\right)$ for chosen values of $\epsilon$
and $a$ in Fig.~\ref{fig:SpasiNSR12-1} and \ref{fig:SpasiNSR12-1-1}.
As the topology of the event horizon is not simple in this geometry
(see \cite{Jo13}) and as it is located in the zone where $r<2\, M$,
we choose to plot only the outer domain for $r\geq2\, M$. The scalar
values are very small compared to the other solutions, which is reassuring
because this example is supposed to be very close to the Kerr spacetime.\\

\section{Conclusion}

We performed the 3+1 decomposition of the Simon-Mars tensor, and defined
two scalar fields from it. It permitted us to quantify the deviation
of different spacetimes to the Kerr one by evaluating these scalars.
This classification works only for stationary spacetimes, but can
be applied to non vacuum spacetimes, and especially in numerical spacetimes
with a matter content, what could not be done before, as far as we
know. Nevertheless, in some of these non-vacuum spacetimes, the scalar
fields can be identically zero even if the spacetime is not locally
isometric to the Kerr spacetime, because the Mars theorem does not
hold in non-vacuum spacetimes. Thus one has to be careful using this
characterization of Kerr spacetime, which provides an efficient way
to compare different spacetimes to one another.

\section*{Acknowledgements}

We thank Jean-Philippe Nicolas for helpful discussions, enlightening
comments and suggestions, Alexandre Le Tiec for interesting and useful
suggestions, and Juan Antonio Valiente Kroon for his help on technical
points.

\end{document}